\documentclass[reprint,amsmath,amssymb,aps,prb,nofootinbib,twocolumn,superscriptaddress,showkeys]{revtex4-2} %

\usepackage{graphicx}
\usepackage{amsthm,amssymb,amsmath,braket,mathdots}
\usepackage{bm}
\usepackage[pagebackref=false,pdfnewwindow=true]{hyperref} 
\usepackage{epstopdf,psfrag}
\usepackage{relsize,amsbsy}
\usepackage[export]{adjustbox}

\usepackage{graphicx,xcolor} 
\usepackage{tabularx, booktabs} 

\usepackage{units} 
\usepackage{dsfont} 

\usepackage{rotating} 

\newcommand{\e}{{\mathrm{e}}} 
\renewcommand{\v}[1]{\bm{#1}} 

\renewcommand{\d}{\mathrm{d}}

\usepackage{tikz}
\usetikzlibrary{arrows.meta}

\definecolor{bananayellow}{rgb}{1.0, 0.88, 0.21}
\definecolor{straw}{rgb}{0.32, 0.28, 0.1}

\def\cblue{\color{black}} 

\usepackage[caption=false]{subfig}
\usepackage{changes}
\definechangesauthor[name=Valentin, color=blue]{VL}
\definechangesauthor[name=Nico, color=red]{NH} 
\definechangesauthor[name=Marc, color=olive]{MAW}

\begin{document}

\title{A Field Guide to non-Onsager Quantum Oscillations in Metals}
\author{Valentin Leeb}
\affiliation{Technical University of Munich, TUM School of Natural Sciences, Physics Department, 85748 Garching, Germany}
\affiliation{Munich Center for Quantum Science and Technology (MCQST), Schellingstr. 4, 80799 M{\"u}nchen, Germany}
\author{Nico Huber}
\affiliation{Technical University of Munich, TUM School of Natural Sciences, Physics Department, 85748 Garching, Germany}
\author{Christian Pfleiderer}
\affiliation{Technical University of Munich, TUM School of Natural Sciences, Physics Department, 85748 Garching, Germany}
\affiliation{Munich Center for Quantum Science and Technology (MCQST), Schellingstr. 4, 80799 M{\"u}nchen, Germany}
\affiliation{Centre for Quantum Engineering (ZQE), Technical University of Munich, 85748 Garching, Germany}
\affiliation{Heinz Maier-Leibnitz Zentrum (MLZ), Technical University of Munich, D-85748 Garching, Germany}
\author{Johannes Knolle}
\affiliation{Technical University of Munich, TUM School of Natural Sciences, Physics Department, 85748 Garching, Germany}
\affiliation{Munich Center for Quantum Science and Technology (MCQST), Schellingstr. 4, 80799 M{\"u}nchen, Germany}
\affiliation{\small Blackett Laboratory, Imperial College London, London SW7 2AZ, United Kingdom}
\author{Marc A. Wilde}
\affiliation{Technical University of Munich, TUM School of Natural Sciences, Physics Department, 85748 Garching, Germany}
\affiliation{Centre for Quantum Engineering (ZQE), Technical University of Munich, 85748 Garching, Germany}

\date{\today} 		
\begin{abstract}
Quantum oscillation (QO) measurements constitute a powerful method to measure the Fermi surface (FS) properties of metals. The observation of QOs is usually taken as strong evidence for the existence of extremal cross-sectional areas of the FS according to the famous Onsager relation. Here, we review mechanisms that generate QO frequencies that defy the Onsager relation and discuss material candidates. These include magnetic breakdown, magnetic interaction, chemical potential oscillations, and Stark quantum interference, most of which lead to signals occurring at combinations of ``parent'' Onsager frequencies. A special emphasis is put on the recently discovered mechanism of quasi-particle lifetime oscillations (QPLOs). We aim to provide a field guide that allows, on the one hand, to distinguish such non-Onsager QOs from conventional QOs arising from extremal cross sections and, on the other hand, to distinguish the various non-Onsager mechanisms from each other. We give a practical classification of non-Onsager QOs in terms of the prerequisites for their occurrence and their characteristics. We show that, in particular, the recently discovered QPLOs may pose significant challenges for the interpretation of QO spectra, as they may occur quite generically as frequency differences in multi-orbit systems, \emph{without} the necessity of visible ``parent'' frequencies in the spectrum, owing to a strongly suppressed temperature dephasing of QPLOs. We present an extensive list of material candidates where QPLOs may represent an alternative explanation for the observation of unexpected QO frequencies.
\end{abstract}

	
\maketitle

\section{Introduction}

The Fermi surface (FS) properties of metals are at the heart of their electronic response functions and have been a focus of scientific investigations since the dawn of band structure theory \cite{Springford1980,kittel2018introduction}.
Quantum oscillations (QOs), discovered in bismuth by de Haas and van Alphen (dHvA) in 1930~\cite{deHaas1930}, have quickly evolved into one of the most important tools to determine FS properties, opening a large and thriving research field dubbed ``fermiology" \cite{Springford1980,1984_Shoenberg_Book,Alexandradinata2023}. 
This tremendous success is rooted in an unrivaled sensitivity to even the smallest FS features and the ability to resolve quasiparticle masses in an orbit-specific way. With the powerful yet simple-to-use Lifshitz--Kosevich formalism \cite{Lifshitz1956}, allowing for analytical data treatment, and the widespread availability of \emph{ab initio} electronic structure calculations \cite{burke2012dft,lejaeghere2016DFT}, amenable to direct comparison, QO experiments are a pillar of experimental FS investigations.

The long list of achievements in fermiology starts out with its contributions to the understanding of electronic transport in the elemental metals \cite{shoenberg1962elemental,Pippard1989}. Up to today, essentially all aspects of modern condensed matter physics concerned with the electronic structure of conductive materials have profited from QO experiments, including conventional and unconventional superconductivity \cite{Yelland2002,Sebastian2015}, heavy-fermion materials \cite{Lonzarich1988,taillefer1988heavy} and itinerant magnets \cite{Lonzarich1980,sebastian2008quantum}.
The basic theory of QO developed for free electrons carries over to correlated quasiparticles~\cite{wasserman1996influence} making QOs the method of choice to verify experimentally if and how electronic correlations or coupling to bosonic quasiparticles renormalize the electronic properties or emergent orders such as charge and spin density or strain waves leading to a reconstruction of the FS \cite{Lonzarich1988,Sebastian2011}.
Even in uncorrelated metals without spontaneous symmetry breaking, electronic properties are often decisively influenced by the exact band placement with respect to the Fermi level, which is still not within the reach of the predictive power of electronic structure calculations. The class of topological semimetals may serve as an excellent example here \cite{gao2019topological,Alexandradinata2023,robredo2024multifold}.

In a typical QO experiment a material exhibits periodic QOs as a function of the inverse applied magnetic field. Following the canonical theory of Onsager~\cite{1952_Onsager_null}, and explained in great detail in the book by Shoenberg~\cite{1984_Shoenberg_Book}, the frequencies are assumed to correspond to the extremal cross-sectional areas of the FS. The damping of the oscillation amplitude with increasing temperature is governed by the mass of the quasiparticle orbiting around the cross section. 

When unexpected extremal orbits are observed experimentally, FS reconstruction or band renormalization are commonly invoked for the interpretation, having important consequences for the understanding of the electronic structure of the material. Because of this, it is of utmost importance to distinguish QO frequencies corresponding to extremal FS cross sections from apparent frequencies in the experimental spectra that may arise from other effects. Various such {\it non-Onsager} mechanisms are well-known in fermiology, namely magnetic breakdown (MB), magnetic interaction (MI), Stark quantum interference (QI) \cite{1984_Shoenberg_Book} and chemical potential oscillations (CPOs) \cite{Alexandrov1996,Shepherd1999}.

Recently, another frequency-generating mechanism, based on quantum oscillations of the quasiparticle lifetime (QPL), mediated by interorbit scattering, was reported in the topological-network semimetal CoSi \cite{huber2023quantum}. The mechanism is predicted to be in principle generic for any metal with multiple extremal orbits \cite{huber2023quantum,leeb2023theory}. An important characteristic of these quasiparticle lifetime oscillations (QPLOs) is that they may produce signals at frequencies corresponding to the difference of two Onsager orbits. Remarkably, these frequencies experience a strongly suppressed temperature dephasing governed by the difference of the masses of the Onsager orbits. This raises the question of whether QPLOs have been observed before in the literature but erroneously attributed to FS cross sections, with potentially dire consequences for our understanding of the electronic structure of the materials. Importantly, QPLOs may yield novel information about microscopic scattering processes, because they are sensitive to interorbit and interband scattering probabilities.

In this paper, we first briefly review the distinct mechanisms to obtain non-Onsager frequencies and present a practical classification in terms of the preconditions required for their occurrence. We discuss their most important characteristics, including allowed frequency combinations and temperature dependence, with the aim to serve as a field guide for QO analysis. Secondly, we identify several examples in the literature where QPLOs may represent an alternative explanation for the experimental observation of unexpected QO frequencies. 

The insight that effective interorbit scattering can lead to a sizable temperature-stable difference frequency in the SdH effect has been known for two-dimensional electron gases (2DEGs) for over three decades~\cite{Polyanovsky1988,Raikh1994,Averkiev2001} under the name of magneto-intersubband oscillations (MISOs). Furthermore, similar effects are known to appear in quasi-2D layered parabolic metals \cite{Polyanosky1993,grigoriev2003theory,thomas2008shubnikov,mogilyuk2018magnetic}, where impurities couple the different layers. 2D and quasi-2D systems, where nearly equal effective masses have led to the observation of a temperature-stable difference frequency, include GaAs heterostructures \cite{Coleridge1990,Leadley1992,Goran2009,Leadley1989,Sander1996,Minkov2020}, metals with bilayer crystal structure \cite{Grigoriev2016,Grigoriev2017}, organic sheet metals \cite{Kartsovnik2002} and recently twisted bilayer graphene~\cite{phinney2021strong}. The insight that similar effects in the form of QPLOs are not restricted to (quasi-)2D materials was put forward only recently~\cite{leeb2023theory,huber2023quantum}. Hence, in the present work, we will focus on such bulk systems. Further, we note that this review is not concerned with the rare and exotic cases of anomalous QOs that have been discovered during the last decade, including the observation of anomalous QOs in bulk insulators~\cite{tan2015unconventional,hartstein2018fermi,liu2018fermi,xiang2018quantum,hartstein2020intrinsic} and heterostructures~\cite{xiao2019anomalous,han2019anomalous,wang2021landau,leeb2021anomalous}, which, in turn, led to a flurry of new theoretical proposals beyond the standard LK theory~\cite{knolle2017anomalous,shen2018quantum,erten2016kondo,zhang2016quantum,sodemann2018quantum,lee2021quantum,he2021quantum,leeb2021anomalous,allocca2022quantum,cooper2023quantum,allocca2023fluctuation}.

Our paper is organized as follows. In section \ref{sec:qm}, we review the quantum mechanical origin of QOs by applying Landau quantization to a simple but non-trivial band model featuring two extremal FS cross sections. We discuss the numerical results to set the stage for the following discussion. In section \ref{sec:LK}, we introduce the semiclassical account of QOs and the Lifshitz--Kosevich (LK) formalism commonly used to analyze experimental results. Both these sections will serve as a point of reference for the discussion of non-Onsager mechanisms.
We review the mechanism of QPLOs and the other -- well-known -- mechanisms capable of generating apparent frequencies in the QO spectra in sections \ref{sec:QPLOintro} and \ref{sec:known-mechanisms}, respectively. Finally, candidate materials for QPLOs in the literature are identified and discussed in section \ref{sec:materials}. Some conclusions are drawn in section \ref{sec:conclusion}.

\section{Quantum mechanical description of \texorpdfstring{{QO\MakeLowercase{s}}}{QOs} in a metal}\label{sec:qm}

\begin{figure*}
  \includegraphics[width=\textwidth]{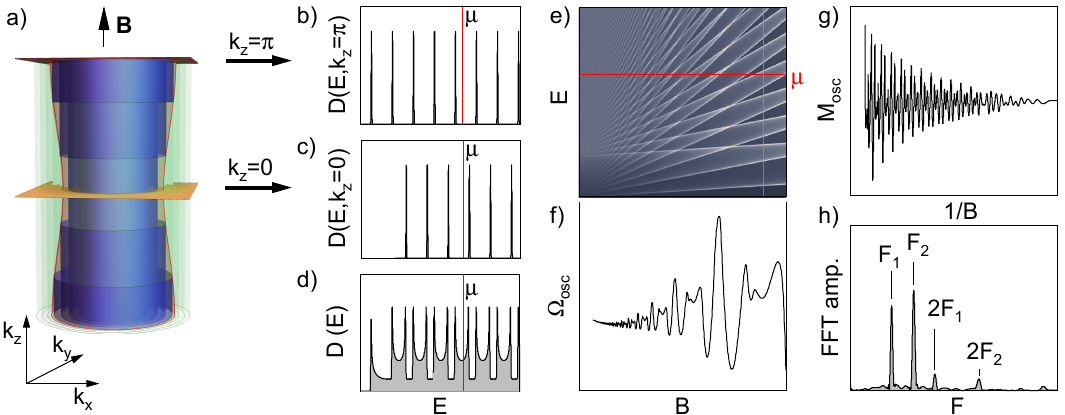}
  \caption{Landau quantization and resulting QOs for a Fermi surface with the shape of a warped cylinder, typical for layered systems with weak interlayer coupling. a) Occupied (blue) and unoccupied (green) sections of the Landau tubes in a magnetic field $\v{B}$ are separated by the FS (red) for $T\rightarrow 0$. 
  b) DOS at $k_z=\pi$. c) DOS at $k_z=0$. d) $k_z$-integrated total DOS. Red lines indicate the position of the chemical potential $\mu$. e) DOS (color scale) versus $E$ and $\v{B}$. The green line indicates the field value used in (a-d). f) Oscillatory part of the grand canonical potential $\Omega_{\rm{osc}}$ exhibiting QOs. g) QOs of the magnetization $\v{M}$ (dHvA effect) as a function of inverse magnetic field. h) FFT of the data in panel (g). Note that distinct frequency peaks occur only at values corresponding to the extremal cross sections of the FS and their higher harmonics. No peaks occur at combination frequencies.}
  \label{fig:warpedcylinder}
\end{figure*}

QOs arise from Landau quantization \cite{landau1930diamagnetismus} of the electronic states in a strong applied magnetic field $\v{B}$. To illustrate the basic mechanism, we assume a band dispersion $\epsilon_{\v{k}}=\frac{\hbar^2}{2m}\left(k_x^2+ k_y^2\right)+t\cos{k_z}$, where $m$ is the effective mass of the quasiparticles in the plane perpendicular to $\v{B}$, and $t$ is a hopping parameter in $z$-direction. The corresponding FS, i.e., the isosurface $\epsilon_{\v{k}}=\mu$, where $\mu$ is the chemical potential, shown in Fig.\,\ref{fig:warpedcylinder}(a) has the shape of a warped cylinder (red). We choose this simple but nontrivial form with two extremal cross sections to facilitate the discussion below and because it resembles a situation frequently encountered in real layered systems with weak interlayer coupling like the dellafossites~\cite{mackenzie2017properties}. An applied magnetic field, taken to point in the $z$-direction, is incorporated via  Peierls' substitution~\cite{peierls1933theorie}  $\v{k}\rightarrow \v{k}+e\v{A}/\hbar$, and restricts the allowed states to lie on circles in any $k_xk_y$-plane with radius $\v{k}_{\perp}=[\frac{2m}{\hbar^2}\hbar \omega_c \left(j+\nicefrac{1}{2}\right) ]^{\nicefrac{1}{2}}$, where $\omega_c=\frac{eB}{m}$ is the cyclotron frequency and $j=0,1,2\dots$ is the Landau quantum number, while it leaves the $k_z$-dispersion unaffected. 
 The states occupied at $T=0$ are colored in blue, while the empty states are shown in green.

The density of states (DOS) $D(E)$ for the $k_z=\pi$ and $k_z=0$ planes shown in Fig.\,\ref{fig:warpedcylinder}(b) and (c), respectively, consists of a series of highly degenerate Landau levels (LLs) at energies $E_j(k_z)=\hbar \omega_c(j+\nicefrac{1}{2})+t\cos{k_z}$. By integrating over $k_z$, the total DOS in Fig.\,\ref{fig:warpedcylinder}(d) is obtained. The total DOS exhibits two series of sharp cusps with smooth tails arising from the $k_z$-integration. The sharp cusps are precisely at the LL positions for $k_z=0$ and $k_z=\pi$, where $\cos{k_z}$ is stationary, while the tails on the high (low) energy side stem from the LLs of different $k_z$-planes that are higher (lower) in energy when the stationary point is a minimum (maximum) of $\cos{k_z}$.

The characteristic shape of each DOS tail results from the $k_z$-dispersion within a Landau tube with given $j$. Where the slope of $\cos{k_z}$ is larger, the Landau levels from adjacent slices at constant $k_z$ are further apart, leading to weaker tails of the DOS. Generalizing this observation, the variation in the DOS, and consequently the amplitude of QOs becomes large when the $k_z$ dispersion is small and vice versa. The tails of the DOS being located on the high (low) energy side of the LL energy that corresponds to the extremal slice gives rise to the well-known phase factor $-(+)\pi/4$ of QOs arising from a maximal (minimal) extremal orbit.

QOs occur when the magnetic field (or the chemical potential) is varied such that the cusps of the DOS and the chemical potential $\mu$ move relative to each other. This simple calculation of the shape of the DOS illustrates why only oscillations are observed that correspond to extremal cross sections (they produce the cusps) while the oscillations due to all other cross sections are smeared out (they do play a role in defining the oscillation amplitudes and the waveform and thus the harmonic content, however). Fig.\,\ref{fig:warpedcylinder}(e) shows $D(E, B)$ as a color plot. For large $j$ and not strictly two-dimensional dispersion, it is often sufficient to assume that $\mu$ remains fixed at its zero-field value when $B$ is varied, as depicted by the red line in (e). When this approximation breaks down, e.g., for low $j$ or two-dimensional Fermi surfaces, the chemical potential oscillations (CPOs) will produce combination frequencies in the spectra as discussed in section \,\ref{sec:known-mechanisms}.

In general, QOs arise in all quantities of a metallic system that depend on the electronic DOS, including all thermodynamic and transport properties. QOs were first reported in the magnetization $M=-\left. \partial \Omega/\partial B \right|_{\mu,T}$, where $\Omega$ is the grand canonical potential, but soon afterward, magneto-transport measurements, i.e., the Shubnikov--de Haas (SdH) effect, were shown to exhibit QOs following the same rules~\cite{Shubnikov1930}.

The properties of many of the oscillatory quantities that have been measured to date are reviewed in the seminal book of Shoenberg \cite{1984_Shoenberg_Book}. Among them, the thermodynamic properties have the advantage that they can be calculated quantitatively without further assumptions. Shown in Fig.\,\ref{fig:warpedcylinder}(f) and \ref{fig:warpedcylinder}(g), respectively, are the oscillatory part, $\Omega_{\mathrm{osc}}(B)$, of the grand canonical potential calculated from
\begin{equation}
    \Omega (\mu,T=0) = \int_{0}^{\mu} (E-\mu) D(E)dE \mbox{ ,}
    \label{eq:omega_t0}
\end{equation}
and the oscillatory part of the magnetization $M_{\mathrm{osc}}(1/B)$ versus inverse magnetic field. A beating of two oscillations is observed. Performing a Fourier Transform of the data yields the frequency spectrum shown in Fig.\,\ref{fig:warpedcylinder}(h). Two strong frequency peaks $F_1$ and $F_2$ are observed together with their weaker harmonics. $F_1$ and $F_2$ correspond to the minimal and maximal extremal cross sections of the FS in Fig.\,\ref{fig:warpedcylinder}(a), respectively, while the harmonics characterize the deviation of the oscillatory waveform from a pure cosine.

The effect of finite temperature can easily be incorporated in the quantum mechanical treatment by considering the Fermi-Dirac distribution $n_{\mathrm{FD}}(E)=1/[ 1+\exp{(E/k_{\mathrm{B}} T  )}] $ of occupied states resulting in~\cite{1984_Shoenberg_Book}
\begin{align}
    \Omega (\mu,T) &=-k_{\mathrm{B}}T \int D(E)\ln{\left[ 1+\exp{\left( \frac{\mu-E}{k_{\mathrm{B}}T} \right)}\right]}dE \\
    & = - \int \frac{dn_{\mathrm{FD}}(E-\mu)}{dE}\Omega (E,0)dE \mbox{ .}
    \label{Eq.3ThermPot}
\end{align}
Thus, finite temperatures smear out the QO with increasing $k_BT/\hbar \omega_c$, i.e., with the ratio of the thermal energy and the cyclotron energy. As a result the amplitude decays for increasing temperature with the LK factor specified in the next section.

\section{Semiclassical description and Lifshitz--Kosevich formalism}\label{sec:LK}

The quantum-mechanical treatment of QOs becomes cumbersome when an analytical description of the band dispersion, and its modification by the quantizing field is not available. Because of that, fermiology usually relies on semiclassical arguments, introduced by Onsager \cite{Onsager1952} and extended by LK~\cite{lifshitz1956theory} and Dingle~\cite{dingle1952some} as follows.

Given an arbitrary band $\epsilon_{\v{k}}$, the quasiparticles are considered as wave packets with a velocity given by $\v{v}(\v{k}) = \hbar^{-1} \nabla_{\v{k}} \epsilon_{\v{k}}$ carrying the electron charge $-e$. In a magnetic field $\v{B}$, the Lorentz force acts on the quasiparticles, leading to the equation of motion $\hbar \dot{\v{k}} = -e \v{v}(\v{k}) \times \v{B}$ (For simplicity, we disregard the effects of Berry phases~\cite{Alexandradinata2023}). Since the Lorentz force cannot perform any work on a charged particle, $\epsilon_{\v{k}}$ is a constant of motion in this classical treatment, forcing the quasiparticles to move on isoenergy contours in momentum space.

\begin{figure*}
  \includegraphics[width=\textwidth]{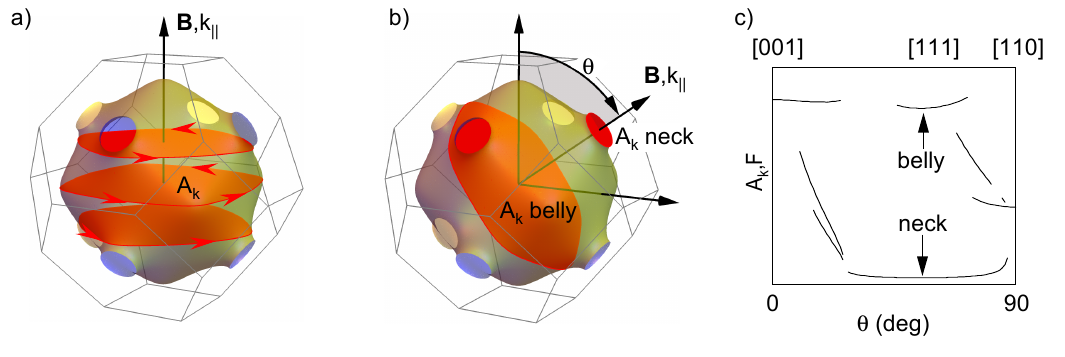}
  \caption{Illustration of the semiclassical QO theory using the FS of elemental copper. Note that only electron orbits are shown for clarity. a) FS of copper (yellow) with extremal cross-sectional areas $A_k$ and quasiparticle orbits (red) for $\v{B}$ along the $z$-axis. b) Extremal cross sections are usually mapped out in several planes of rotation as sketched here for the plane spanned by $[001]$ and $[110]$. c) $A_K$ or $F$ versus angle $\theta$ defined in (b) for the electron orbits of Cu.}
  \label{fig:semiclassical}
\end{figure*}

Following this argument, quasiparticles move on momentum space contours that are given by the intersection of the FS (yellow) with planes perpendicular to the magnetic field as illustrated in Fig.\,\ref{fig:semiclassical}(a) for the FS of Cu. When the contours are closed, these orbits are quantized following the Bohr--Sommerfeld quantization rule $2\pi \hbar (j+\gamma) = \oint \mathbf{p} \mathrm{d}\mathbf{r}$, where $\mathbf{p}$ is the canonical momentum and $\gamma$ is a phase factor arising from multiple contributions including the Berry phase~\cite{Alexandradinata2023}. 
Only orbits with an enclosed area that is extremal with respect to $k_z$ (red) are considered because they determine the QOs as outlined above. Onsager's famous result~\cite{1952_Onsager_null} now states that the momentum space area $A_k$ enclosed determines the QO frequency $F$ arising from the extremal semiclassical orbit according to
\begin{equation}
   F= \frac{\hbar}{2 \pi e} A_k(\mu) \mbox{ .}
   \label{eq:Onsager}
\end{equation}
Eq.~\eqref{eq:Onsager} turns QO experiments into a spectroscopic tool capable of mapping out the extremal FS cross sections as a function of field orientation as sketched in Fig.\,\ref{fig:semiclassical}(b) and (c).

\begin{figure}
  \includegraphics[width=\linewidth]{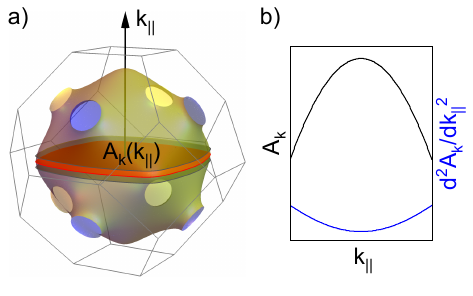}
  \caption{a) Illustration of the curvature factor occurring in the LK-formula, i.e., the $2^{\mathrm{nd}}$ derivative of the area $A_k$ with respect to $k_{\parallel}$. b) $A_k(k_{\parallel})$ (black) and the $2^{\mathrm{nd}}$ derivative (blue). A low curvature corresponds to a large QO amplitude because the oscillations in the DOS are stronger.}
  \label{fig:curvature}
\end{figure}

LK subsequently established an analytical description of QOs in the semiclassical limit that is based on a Fourier series of the oscillating part of the grand canonical potential $\Omega$ in Eq.\,\eqref{eq:omega_t0}. The oscillatory part of the grand canonical potential for a single extremal orbit is given by
\begin{equation}
    \Omega_{\mathrm{osc}}\!=\! \frac{e^2}{4\pi^4m} \sqrt{\frac{2\pi e}{\hbar \frac{\partial^2 A_k}{\partial k_{\parallel}^2}}} \sum_{p=0}^{\infty} \frac{B^{5/2}}{p^{5/2}} \cos{\left[2\pi p \left(\! \frac{F}{B}\!-\!\gamma \!\right)\! \mp \frac{\pi}{4} \right]} \mbox{,}
    \label{eq:omegaosc}
\end{equation}
where $p$ is the harmonic index and $\partial^2 A_k/\partial k_{\parallel}^2$ is the curvature of the reciprocal space area $A_k$ with respect to the $\v{k}$-component parallel to the applied magnetic field, as illustrated in Figs.\,\ref{fig:curvature}(a) and (b) \cite{1984_Shoenberg_Book}. The prefactor representing the curvature can be understood intuitively from the quantum mechanical calculation, because increasingly large portions of a Landau tube pass through the chemical potential simultaneously with decreasing curvature, leading to larger oscillation amplitudes.

{\cblue The phase in the LK formula [Eq.~\eqref{eq:omegaosc}] is set by a contribution $\mp \pi/4$ with $+$ ($-$) for minimal (maximal) orbits arising from the DOS tails of three-dimensional FS sheets and a contribution $\gamma$ containing further phase angles. In the simplest case, $\gamma$ is set by the Berry phase acquired along the semiclassical orbit and the Maslov index, i.e., for a strictly parabolic band with zero Berry phase, $\gamma=1/2$ whereas for Dirac fermions, $\gamma=0$ (e.g. in graphene~\cite{goerbig2011electronic}). In general, $\gamma$ may contain various contributions and is not restricted to discrete values. See, e.g., the extensive discussion in Ref.\,\cite{alexandradinata2018semiclassical} for details. From here on, we will encapsulate all phase contributions in the single variable $\phi_p$. In bulk 3D metals far from the quantum limit, i.e., where $F/B \gg 1$, the absolute phase is difficult to determine reliably in experiments because determining the phase requires an extrapolation to $B \approx F$, naturally inducing large uncertainties. However, the relative phases between QO frequencies can drastically affect the QO amplitudes \cite{mccollam2005anomalous}. In addition, Zeeman splitting of spin degenerate FS sheets may lead to an amplitude suppression and phase shift or even a complete cancellation of all odd (or even) harmonics, called the spin-zero effect \cite{Shoenberg1984}. The effects of linear Zeeman splitting, often incorporated into a spin damping factor, are discussed in detail in Ref.~\cite{Shoenberg1984,Alexandradinata2023}. We neglect the spin damping factor for simplicity.}

In this framework, damping of the QOs due to finite temperature $T$, finite QPL $\tau$, and further mechanisms may be incorporated in a unified way as prefactors via the Fourier transforms of suitable phase distribution functions. Most notably, LK connected the temperature dependence of the amplitude of QOs to the cyclotron effective mass of the quasiparticles~\cite{Lifshitz1956}
\begin{equation}
    m =\frac{\hbar^2}{2 \pi} \frac{\partial A_k(\mu)}{\partial E} \mbox{ .}
    \label{eq:cyclotronmass}
\end{equation}
\begin{figure}
  \includegraphics[width=\linewidth]{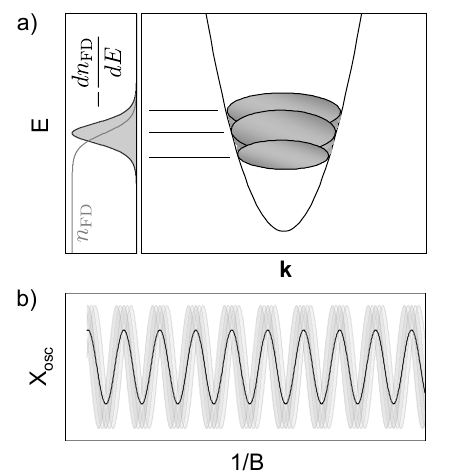}
  \caption{a) Illustration of the cyclotron mass definition. The effect of finite $T$ may be thought of as a distribution of orbits at slightly different energy, given by $-dn_{\mathrm{FD}}/dE$. How the distribution of energies relates to the distribution of cross section scales with $dA_k/dE$. b) QOs in a quantity $X$ arising from superposition of orbits with slightly different $A_k$, or, equivalently, slightly different phase. The strength of the $T$-dephasing scales with $dA_k/dE$.}
  \label{fig:T-dependence}
\end{figure}
This may be understood as follows. The quantum oscillations of a single extremal orbit at $T=0$  as given in Eq.\,\eqref{eq:omegaosc} exhibit a single, well-defined frequency $F$ and phase. Now, QOs at a finite temperature can be thought of as a superposition of oscillations at slightly different energies [with the distribution given by $-dn_{\mathrm{FD}}/d E$, as shown in Fig.\,\ref{fig:T-dependence}(a)] and thus slightly different frequencies $F$ or, equivalently, slightly different phases $\phi$ 
as sketched in Fig.\,\ref{fig:T-dependence}(b). The strength of the $T$-damping, depends on how strongly $F\propto A_k$ varies with $\mu$, which is directly reflected in the definition of the cyclotron mass of Eq.\,\eqref{eq:cyclotronmass}, being proportional to the slope of $A_k(\mu)$.

Formally, the phase averaging can be written as a convolution of the oscillatory quantity at zero temperature with a phase distribution function $g(\psi/\lambda)=1/(1+\cosh \psi/\lambda)$ scaled by $\lambda=2\pi p k_BT/\hbar \omega_c$ \cite{1984_Shoenberg_Book}, i.e., for the $p$-th harmonic

\begin{equation}
{\cblue 
   \Omega_{\mathrm{osc},p}\propto \frac{\int_{-\infty}^{\infty} \cos \left[ 2\pi p \frac{F}{B} + \phi_p  +\psi \right] g(\psi/\lambda) \d\psi}{\int_{-\infty}^{\infty} g(\psi/\lambda) \d\psi}  \mbox{ .} 
}
   \label{eq:phase-averaging}
\end{equation}

Eq.~\eqref{eq:phase-averaging} can be transformed into
\begin{equation}
   \Omega_{\mathrm{osc}}\propto \frac{f(\lambda)}{f(0)} \cos \left[ 2\pi p {\cblue \frac{F}{B} + \phi_p}  \right]  \mbox{ ,} 
\end{equation}
where $f(\lambda)/f(0)=\pi \lambda/ \sinh \pi \lambda$ is the normalized Fourier transform of $g(\psi/\lambda)$ with respect to $\lambda$. This represents the so-called LK temperature damping factor
\begin{equation}
   R_T(p m) = \frac{2\pi^2 p m k_\mathrm{B} T/\hbar eB}{\sinh \left(2\pi^2 p m k_\mathrm{B} T/\hbar eB\right)} \mbox{ } 
\end{equation}
resulting from the Fourier transform of $-d n_{\mathrm{FD}}/dE$, see Eq.~\ref{Eq.3ThermPot}. The same approach can be used to account for other dephasing mechanisms due to scattering, like sample mosaicity, magnetic-field inhomogeneity, etc. as long as they are described by symmetric distribution functions~\cite{1984_Shoenberg_Book}.

Impurity scattering broadens the LLs with a half-width proportional to the inverse of the QPL $\tau$~\cite{1984_Shoenberg_Book}. This lifetime is expected to oscillate with the DOS at the Fermi level $D(\mu)$, as both the states available for scattering and screening depend on $D(\mu)$. Because the oscillations in $\tau$ have the same frequency $F$ as the QOs, and thus can only modify the harmonic content in Eq.\,\eqref{eq:omegaosc}, they are usually neglected, and the effects of scattering are described by introducing a phenomenological constant, which is expressed either as a lifetime $\tau$ or an equivalent thermal energy $k_{\mathrm{B}}T_{\mathrm{D}}$ with the so-called Dingle temperature $T_{\mathrm{D}}$. Following the same derivation as for temperature, but assuming a Lorentzian lifetime broadening, following Eq.~\eqref{eq:phase-averaging}, it can be shown~\cite{1984_Shoenberg_Book} that this leads to an exponential suppression of the QOs, known as the Dingle reduction factor 
\begin{equation}
    R_{\mathrm{D}}(p T_{\mathrm{D}}) = \e^{-\frac{2\pi^2 p m k_{\mathrm{B}} T_{\mathrm{D}}}{\hbar eB}} \mbox{ .}
\end{equation}
An increase of $T_{\mathrm{D}}$ leads to a magnetic field-dependent amplitude suppression of the QOs. 

To summarize, in the canonical theory of QOs a given frequency is related to one extremal orbit around the FS perpendicular to the magnetic field. The oscillating part of an observable $X$ may then be described by
\begin{align}
    X_\text{osc} \propto \sum_p \cos\left(2 \pi p \frac{F}{B} {\cblue + \phi_p}\right) R_{\mathrm{T}}(p m) R_{\mathrm{D}}(p T_{\mathrm{D}}) \mbox{ .}
\end{align}
When multiple extremal orbits are present, they appear as a superposition of independent oscillations $X_{\mathrm{osc}}=\sum_i X_{\mathrm{osc},i}$ {\cblue where $i$ indexes frequency $F_i$, effective mass $m_i$, phase $\phi_{p,i}$, and Dingle temperature $T_{D,i}$.}

This canonical theory of Onsager, LK, and Dingle is the foundation of fermiology and has been used to interpret QO experiments on systems ranging from the elementary metals \cite{1984_Shoenberg_Book} over strongly correlated heavy-fermion compounds \cite{1987_Taillefer_JMMM} to cuprate high-temperature superconductors \cite{doiron-leyraud2007quantum}.  In Sec.~\ref{sec:known-mechanisms} we will review known mechanisms and their prerequisites which can lead to additional QO frequencies. We stress that as long as these can be excluded the standard FS determination using the LK theory has proven to be extremely successful for many decades.

The QPLO mechanism proposed recently is a new player in the game. It is generic in the sense that it may be operational for any metallic system that exhibits more than one QO orbit with some nonlinear coupling of the orbits.

\section{Quasiparticle Lifetime Oscillations}\label{sec:QPLOintro}

In the following, we briefly describe the mechanism leading to QPLOs.
\begin{figure*}
  \includegraphics[width=\textwidth]{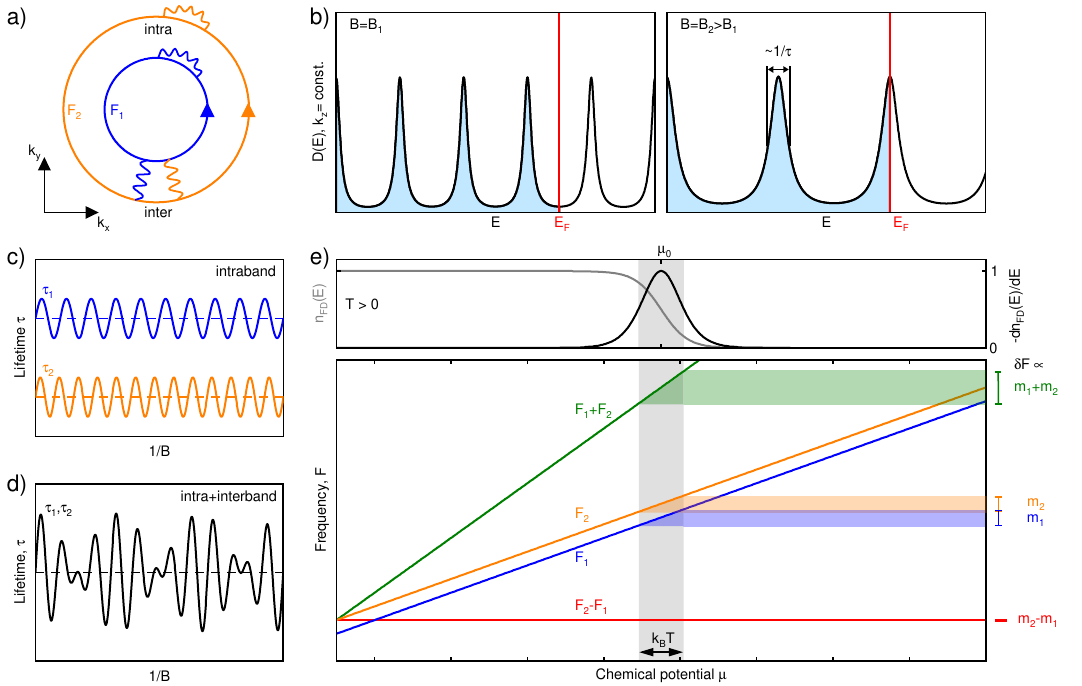}
  \caption{Origin of the QPLO. a) Depiction of two FS orbits with QO frequencies $F_1$ and $F_2$. Intraband scattering causes transitions within each FS orbit; interband scattering causes transitions between the FS orbits. The scattering effectively mixes the orbits. b) DOS $D(E)$ due to Landau quantization of the cyclotron orbits under magnetic field $B_1$ and $B_2>B_1$. The Landau-level separation and degeneracy varies linearly with $B$ such that the DOS at the Fermi level is an oscillatory function of $1/B$. Scattering causes a finite lifetime broadening of the Landau levels. Because the lifetime $\tau$ of the cyclotron orbits varies with $D(\mu)$, it is also an oscillatory function of $1/B$. c) Oscillatory QPL associated with FS orbits $1$ and $2$ with intraorbit but no interorbit scattering. d) Oscillatory QPL associated with FS orbits $1$ and $2$ in the presence of intraorbit and interorbit scattering. As the interorbit scattering depends on the DOS on both FS orbits $1$ and $2$, the QPL oscillates with a superposition of the DOS of both orbits. e) Connection between the temperature dependence of the oscillation amplitude and the effective mass. A parabolic band dispersion results in a linear frequency (area) versus chemical potential relationship $F(\mu)$. Two bands with parallel dispersions $F_{1,2}(\mu)$ lead to a complete suppression (doubling) of the dispersion for oscillations at $F_2-F_1$ ($F_1+F_2$). The temperature dependence of QOs is governed by the variation of frequency $\delta F$ in an interval $k_BT$ around $\mu_0$ depicted in gray shading and given by the negative derivative of the Fermi distribution with respect to energy, $-dn_{FD}(E)/dE$. 
  Reproduced with permission.\textsuperscript{\cite{huber2023quantum}} \citeyear{huber2023quantum}, Springer Nature. }
  \label{fig:schematicIdea}
\end{figure*}
Given a generic metal in 2D or 3D with at least two extremal FS cross sections $A_{k1}$ and $A_{k2}$, one expects according to the canonical Onsager theory one QO frequency ($F_{1,2,\dots}$ ) for each extremal semiclassical orbit and higher harmonics thereof. The key ingredient for the appearance of QPLOs is a non-linear coupling of orbits. Most natural are different kinds of scattering between the bands. Here, we restrict the discussion to defect scattering as sketched in Fig.\,\ref{fig:schematicIdea}(a) and analyzed in Ref.~\cite{leeb2023theory}.

It is important to differentiate between the usual intraorbit scattering and \emph{interorbit} channels, which allow quasiparticles to scatter between two different extremal orbits that may belong to the same or a different band. For the latter, interorbit and interband are the same.

As explained above, defect scattering causes a broadening of the Landau levels, in which the inverse of the QPL, $1/\tau$, corresponds to the half-width of the peaks in the DOS. Assuming the Born approximation and Fermi’s golden rule, the QPL varies with the DOS at the Fermi energy, as sketched in Fig.\,\ref{fig:schematicIdea}(b) for two different magnetic field values. Hence, as a function of the inverse magnetic field, the lifetimes $\tau_{\mathrm{1}}$ and  $\tau_{\mathrm{2}}$ of two orbits include a periodic oscillatory component as well as their average value as shown in Fig.\,\ref{fig:schematicIdea}(c).

In the presence of intraorbit scattering, the oscillatory component of $\tau$ varies with the QO frequency $F_1$ of the underlying FS orbit. Remarkably, in the presence of intraorbit and interorbit scattering to a second orbit with $F_2$, 
 the lifetime $\tau$ oscillates with {\it both} frequencies of the participating FS orbits, $F_1$ and $F_2$. Expressed in terms of the Dingle temperature associated with the QPL, $T_{\mathrm{D}}$, and the Dingle temperature of the non-oscillatory average, $T_{\mathrm{D0}}$, the inverse of the QPL may be written as
\begin{equation}
    \frac{1}{\tau} \propto T_{\mathrm{D}} \propto T_{\mathrm{D0}} + A_1 \cos \left(2\pi\frac{F_1}{B} \right) + A_2 \cos \left(2\pi\frac{F_2}{B} \right) \mbox{ ,}
    \label{eq:QPLT}
\end{equation}
where $A_1$ and $A_2$ are prefactors. {\cblue Here and in the following, we omit the phases for simplicity.}

In physical quantities, the QPLOs cause further oscillatory components expressed conveniently in terms of $T_{\mathrm{D}}$. Owing to an additional linear dependence on $T_{\mathrm{D}}$, the by far strongest effect is observable in electrical-transport properties like the electrical conductivity
\begin{equation}
    \sigma \propto T_{\mathrm{D}} \left[ \cos \left(2\pi\frac{F_1}{B} \right) + \cos \left(2\pi\frac{F_2}{B} \right) \right] R_{\mathrm{D}}(T_{\mathrm{D}}).
\end{equation}

Inserting $T_{\mathrm{D}}$ from Eq.\,\eqref{eq:QPLT} one obtains to leading order terms in \mbox{$\cos\left( 2\pi F_{1}/B\right)\times\cos( 2\pi F_2/B)$} and $\cos^2( 2\pi F_\lambda/B)$ ($\lambda=1,2$). Using trigonometric identities, the former decomposes into $\cos(2\pi (F_2-F_1)/B)$ and $\cos(2\pi (F_2+F_1)/B)$. Note that there is a fixed phase relation $\phi_\pm = \phi_2 \pm \phi_1$, which we did not include for simplicity. Further subleading oscillatory dependencies can be generated as well, such as $2 F_1-F_2$, $2 F_2 -F_1$, $2 F_2- 2 F_1$, $3 F_2 - F_1$, and so forth. As for the terms in $\cos^2( 2\pi F_\lambda/B)$ ($\lambda=1,2$), these will only modify the higher harmonics of the fundamental frequencies.

In contrast, different behavior occurs for the magnetization 
\begin{equation}
    M_\text{osc} \propto \left[ \cos \left(2\pi\frac{F_1}{B} \right) + \cos \left(2\pi\frac{F_2}{B} \right) \right] R_{\mathrm{D}}(T_{\mathrm{D}}) \mbox{ ,}
\end{equation}
where the additional factor of $T_{\mathrm{D}}$ is absent. Here all difference frequency combinations cancel exactly for impurity scattering, whereas sum combinations remain~\cite{leeb2023theory}. 

Apart from the emergence of a new difference frequency the most striking aspect of QPLOs concerns the temperature dependence. The fundamental QO frequencies are determined by the extremal cross-sectional areas $A_{\textrm{k}}(\mu)$, which in general vary as a function of the chemical potential $\mu$. Accordingly, oscillations at $F_2 \pm F_1$ are also expected to change as a function of  $\mu$, see  Fig.\,\ref{fig:schematicIdea}(e). Recalling that $R_{\rm T}(T)$ originates from an averaging of orbits with different $\mu$, weighted by $-dn_{\mathrm{FD}}/dE$, one finds that the width of the distribution of Fermi levels at finite temperature, $\delta \mu \sim k_{\mathrm{B}}T$, causes a distribution of frequencies, $\delta F$, that scales with the cyclotron mass, $\delta F/k_{\mathrm{B}}T=\hbar/(2\pi e)\, \partial A_{\textrm{k}}/\partial E = m/(e\hbar)$. Consequently, for the oscillations at $F_2 \pm F_1$ finite temperatures cause distributions given by $\delta (F_2 \pm F_1)/k_{\mathrm{B}}T=\hbar/(2\pi e) \,\partial (A_{\textrm{k},2}\pm A_{\textrm{k},1})/\partial E =(m_2\pm m_1)/(e\hbar)$.

It follows that a difference frequency $F_2-F_1$ can occur in the experimental spectra that is insensitive to temperature when the cyclotron masses $m_2$ and $m_1$ are similar. Such difference frequencies can even persist in a temperature regime where the ``parent''
 oscillations have already vanished, making it potentially hard to identify the oscillation as a difference frequency in the first place.

The discussion above applies to two-electron orbits coupled by interorbit impurity scattering. It can equivalently be applied to two hole orbits. The situation is different when considering an electron orbit coupled to a hole orbit. Since a hole orbit can be assigned a negative cyclotron mass, the sum frequency obtains an effective mass $||m_\text{particle}| - |m_\text{hole}||$, and can potentially become temperature stable, whereas the difference frequency is now strongly temperature dependent scaling with $|m_\text{particle}| + |m_\text{hole}|$. It is worthwhile to compare this with a simple case of magnetic breakdown (MB) (see next sec.~\ref{sec:MB}). For MB, between two electron orbits or two hole orbits, only sum frequency combinations are allowed because otherwise, the semiclassical trajectory of the quasiparticle would be partially in opposition to the Lorentz force. However, MB leading to a mixed electron-hole orbit will result in a difference frequency, whereas now the sum frequency is semiclassically forbidden. In both cases, the $T$-dependence of the MB combination frequencies is governed by the sum of the effective masses, in contrast to QPLO frequencies.

We note that the QPLO is a second-order scattering process and, therefore, the magnetic field dependence is governed by the product of the Dingle factors of each orbit~\cite{leeb2023theory}. 

{\cblue The relative amplitudes of QPLOs and fundamental frequencies is set by the strength of interorbit scattering between the fundamental orbits. The scattering matrix elements are material-specific, e.g., via the orbital content of the participating trajectories and sample-dependent via the type and concentration of scattering centers. Therefore, the observation of QPLOs has been proposed as a versatile tool to perform impurity spectroscopy \cite{leeb2024numerical,leeb2024interband}.
}

Following these considerations, QPLOs are not only a novel mechanism generating non-Onsager QO frequencies but may also yield information about quasiparticle scattering in a wide range of material classes.

\section{Conventional mechanisms generating non-standard QO frequencies}\label{sec:known-mechanisms}

When considering experimental QO spectra, one needs to distinguish standard Onsager QO from various non-Onsager mechanisms, which can generate sharp frequency peaks that do not correspond to fundamental extremal FS cross sections or integer multiples thereof. 
\begin{figure*}
  \includegraphics[width=0.85\textwidth]{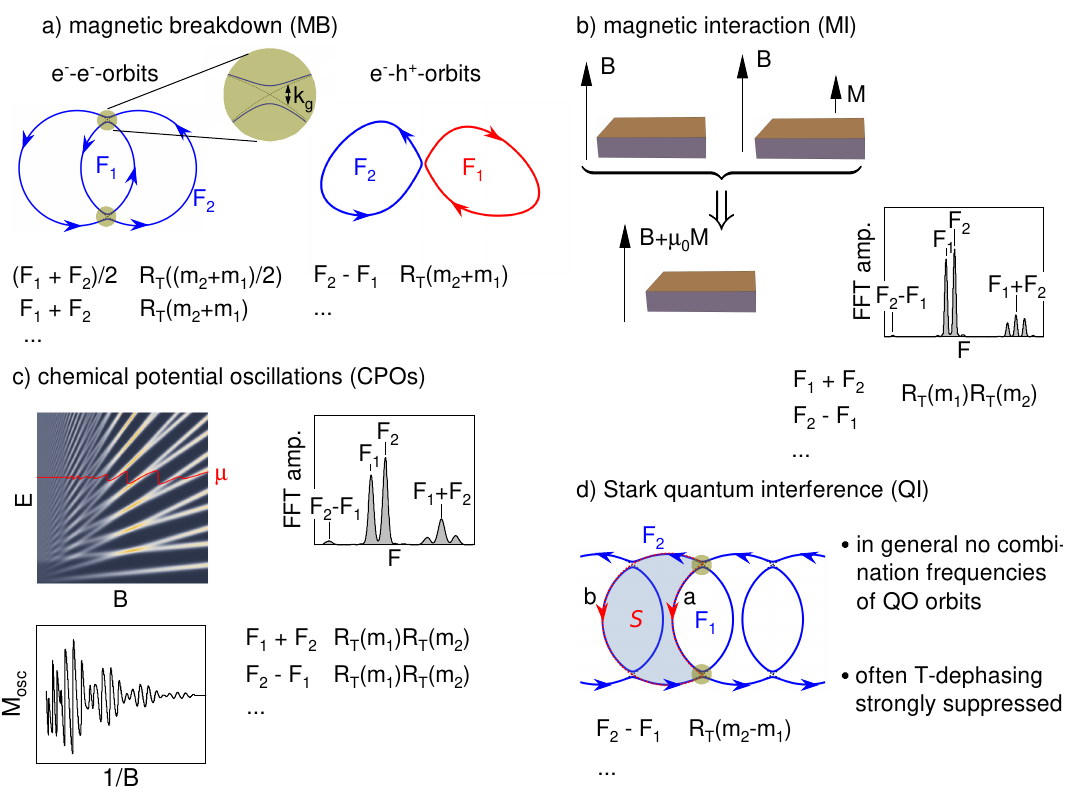}
  \caption{Overview over conventional mechanisms that may lead to apparent QO frequencies that do not correspond to fundamental extremal orbits and corresponding minimal examples a) Magnetic breakdown. Two intersecting electron orbits and a combined electron-hole orbit coupled by magnetic field-assisted tunneling at the breakdown junctions. In general, extremely rich combination frequency spectra can be generated by MB. Difference frequencies may occur only for e-h-orbits. The strong $T$-dephasing is governed by the sum of the individual orbit masses. b) Magnetic interaction. The backaction of $M$ on the internal field $B$ distorts the field axis and leads to combination frequencies including differences. $T$-dephasing scales with the product of the individual LK-factors. FFT for the case of MI of two equally strong QOs at $F_1$ and $F_2$, see text. c) Chemical potential oscillations. For a 2DES with two occupied subbands and low carrier density, the chemical potential (red line) oscillates strongly. The FFT of $M_{osc}(1/B)$ exhibits combination frequencies including differences. $T$-dephasing scales with the product of the individual LK-factors. d) Stark quantum interference. For chainlike trajectories as they may arise from a 1D modulation with finite $q$-vector, partial magnetic breakdown may lead to an apparent difference frequency with a $T$-dephasing governed by the difference of the associated masses. {\cblue Here, $F_1$ is the lentil-shaped fundamental FS orbit, whereas the circular $F_2$-orbit is enabled by MB.}. The interferometer {\cblue area $\mathcal{S}$, shaded in blue, is defined by the two pathways a and b shown as red dashed lines}. In general, QI oscillations arise in kinetic quantities from the interference of coherent quasiparticle pathways in effective interferometer configurations mediated by partial magnetic breakdown. QI does not generally lead to frequencies that are any combination frequencies of fundamental QO orbits. The $T$-dephasing is often strongly suppressed, because QI pathways with the same direction of QP motion tend to be shifted in reciprocal space to first order upon variation of $\mu$, while a change of the interferometer area appears only in second order.}
  \label{fig:coventionalMech}
\end{figure*}
In the following, we briefly summarize the well-established conventional mechanisms and the conditions for their occurrence with the intention to provide a compact overview of the different characteristics on the one hand and to distinguish them from QPLOs on the other hand.

\subsection{Magnetic breakdown (MB)}
\label{sec:MB}
Magnetic breakdown (MB) represents field-assisted tunneling of electrons between adjacent orbits in reciprocal space \cite{Cohen1961,Blount1962,1966_Chambers_ProcPhysSoc,slutskin1968dynamics} as schematically shown in the left panel of Fig.\,\ref{fig:coventionalMech}(a) for two electron orbits. The conditions for the occurrence of MB are quite stringent, because the probability $p$ for an interband tunneling event depends exponentially on the breakdown field $B_0$, i.e., $p=e^{-B_0/B}$, where $B_0$ may be conveniently expressed in terms of the geometric orbit properties by the Chamber's formula \cite{1966_Chambers_ProcPhysSoc}
\begin{equation} 
    B_0=\frac{\pi \hbar}{2e}\sqrt{\frac{k_g^3}{a+b}} \mbox{ .}
    \label{eq:breakdown-field}
\end{equation}
 Here, $k_g$ is the gap in reciprocal space and $a$ and $b$ are the curvatures of the two orbits at the junction. Thus, MB may occur only when different parts of the FS are separated by very small gaps, for example arising from Bragg reflection on weak periodic potentials, from the lifting of band degeneracies by weak spin-orbit coupling, or backfolding of density wave orders.

MB between only electron orbits or between only hole orbits can lead to various integer combination frequencies, with effective masses being the sum of the absolutes of the effective masses of the contributing orbits, such that the temperature dependence of an MB frequency $F_1+F_2$ is governed by $R_{\mathrm{T}}(m_1+m_2)$. Therefore, they are strongly damped in temperature \cite{1984_Shoenberg_Book}. Due to the vast possibility of combinations the frequency spectra can be --- depending on the geometry of the FS and the number of breakdown junctions --- extremely complicated \cite{1983_Kaganov_PhysicsReports}. The occurrence of frequencies at the difference frequency of fundamental orbits is semiclassically forbidden, as illustrated in Fig.\,\ref{fig:coventionalMech}(a), because an electron on an orbit enclosing $(F_2-F_1)/2$ would have to move in opposition to the Lorentz force on part of the orbit.

A case in which MB can lead to frequencies corresponding to the difference of two basis frequencies is a mixed electron-hole orbit~\cite{obrien2016magnetic} as illustrated by the figure-of-eight path in the right panel of Fig.\,\ref{fig:coventionalMech}(a), which is semiclassically allowed. Here, the observed breakdown frequency is $|F_2-F_1|=(\hbar/(2\pi e))| A_{\textrm{k},2}- A_{\textrm{k},1}|$, because the areas enter into the formula with different signs. 

This may be understood intuitively in the semiclassical picture: In the Bohr-Sommerfeld quantization condition, $\oint \mathbf{p} \mathrm{d}\mathbf{r}$ evaluates effectively to the {\cblue signed} area enclosed by the semiclassical trajectory of the quasiparticle. The sign of this expression depends on the sense of rotation of the quasiparticle trajectory. However, even in {\cblue the case of mixed electron-hole orbits} the difference frequency is strongly temperature damped. {\cblue The absolute area of the hole orbit decreases with increasing energy whereas the absolute area of the electron orbit increases. Thus,} the signs of $\partial A_{\textrm{k},2}/\partial E$ and $\partial A_{\textrm{k},1}/\partial E$ are {\cblue opposite, and hence $| A_{\textrm{k},2}- A_{\textrm{k},1}|$ varies strongly with $E$}. The temperature dependence of the electron-hole orbit is again governed by the sum of the absolute of the cyclotron masses of the individual orbits~\cite{obrien2016magnetic}, i.e., $R_{\mathrm{T}}(|m_1|+|m_2|)$.

A hallmark characteristic of MB is the magnetic field dependence arising from the exponential dependence of the tunneling probability on the magnetic field. Because the breakdown fields in Eq.\,\eqref{eq:breakdown-field} can differ vastly for different MB junctions the behavior is in general very complex. However, as a trend, MB tends to be absent in sufficiently low fields, such that only the fundamental frequencies are observed. In a crossover regime, both fundamental and MB frequencies are observed, while in the high-field limit, only the MB orbits survive and the fundamental oscillations are suppressed.

\subsection{Magnetic interaction (MI)}
Magnetic interaction (MI) in the context of additional QO frequency contributions refers to the feedback of QOs of the magnetization $M_\text{osc} $, i.e., the dHvA effect, onto the internal field $B_{\mathrm{i}} \rightarrow B_{\mathrm{a}} + \mu_0 M_\text{osc}$,  where $\mu_0$ is the vacuum permeability and $B_{\mathrm{a}}$ is the applied field, see Fig.\,\ref{fig:coventionalMech}(b).  MI becomes only appreciable, when $\mu_0 dM_\text{osc}/dB \not \ll 1$. In the limit $\mu_0 dM_\text{osc}/dB > 1$, the system becomes unstable and the dHvA effect becomes hysteretic. MI may drastically change the QO spectrum and, for the case of two dHvA frequencies, $M_\text{osc} = A_1 \cos(2\pi F_1/B_{\mathrm{i}}) R_{\mathrm{T}}(m_1) + A_2 \cos(2\pi F_2/B_{\mathrm{i}}) R_{\mathrm{T}}(m_2)$, 
 generate strong frequency components at $F_2\pm F_1$ as illustrated by the FFT spectrum in Fig.\,\ref{fig:coventionalMech}(b) calculated for two equally strong oscillations and $\mu_0 dM_\text{osc}/dB \approx 1/2$. In the regime between the limiting cases, the leading contributing terms are of the form 
\begin{align}
   A_i \cos \! \left( \! 2\pi\frac{F_i}{B + \mu_0 M_\text{osc}} \! \right) \approx&
    A_i \cos \left(2\pi\frac{F_i}{B}\right) + \nonumber \\ & A_i \sin \left(2\pi\frac{F_i}{B}\right) \frac{2\pi F_i}{B^2} \mu_0 M_\text{osc}.
\end{align}
The magnetization oscillates with both frequencies and a prefactor $\sqrt{B}$ as it can be derived from Eq.~\eqref{eq:omegaosc}. Hence, the second summand includes terms of the form $2 \sin \left(2\pi F_1/B\right) \cos \left(2\pi F_2/B\right) = \sin\left(2\pi (F_1-F_2)/B\right) + \sin\left(2\pi (F_1+F_2)/B\right)$ leading to sum and difference frequencies. The amplitudes of the sum and difference frequencies are given by $A_\pm = \frac{\pi (F_1 \pm F_2)}{B^2} \mu_0 A_1 A_2$. It follows that the relative amplitudes of the MI-induced sum and difference frequencies are given by $A_\pm/A_i \approx \mu_0 \d M_\text{osc} / \d B$, recovering the condition stated above \cite{1984_Shoenberg_Book}. The sum and difference oscillation frequencies due to MI vanish quickly with increasing temperature as the product of the reduction factors $R_{\mathrm{T}}(m_1)R_{\mathrm{T}}(m_2)$.

Assuming that the field dependence of the fundamental oscillations is governed solely by the Dingle reduction factor at sufficiently low temperatures, the field dependence of $A_{\pm}$ is non-monotonic: $A_{\pm}$ first rises sharply with the product of the two Dingle factors, passes a maximum and approaches a $1/B^2$ dependence in the high-field limit, where the Dingle factors have saturated.

As a sidenote, another feedback mechanism termed torque interaction may occur in torque magnetometry \cite{1984_Shoenberg_Book,bergemann1999quantum} due to the change in torque with angle occurring in response to the deflection or torsion of the sensor. As this is a purely instrumental effect and can be at least in principle avoided by measuring torque at constant position, we do not discuss this mechanism further.

\subsection{Chemical potential oscillations (CPOs)}

Quantum oscillations of the chemical potential are, in principle, always present in a floating metallic system in a quantizing magnetic field because the conserved quantity is the number of electrons. Any variation in the free energy of the system must be compensated by a change in the chemical potential in order to satisfy charge conservation. In a multiband metal with several Fermi surface pockets, CPOs may lead to a mixing of different oscillation frequencies~\cite{1997_Alexandrov_PhysicsLettersA}. CPOs are most pronounced for a \mbox{(quasi-)2D} Fermi surface, i.e. for almost no $k$-dependence of the energy parallel to the field direction, and with only a few Landau levels occupied, i.e., close to the quantum limit where $F \sim B$~\cite{1984_Shoenberg_Book}. It is illustrated in Fig.\,\ref{fig:coventionalMech}(c), where the Landau level structure and chemical potential, oscillatory magnetization and FFT spectrum for a 2D electron system with two occupied subbands is shown. The CPOs are strong (red line) because of the 2D Landau levels and the low Landau level index, leading to combination frequencies in the FFT spectrum, most prominently at $F_2\pm F_1$. In turn, CPOs are well-known in experimental dHvA spectra of \mbox{(quasi-) 2D} materials.

The origin  of CPOs is a fixed particle number. The particle number $N$ and the chemical potential are related via a self-consistent integral equation $N = \int_{-\infty}^{\mu(B)} D(E,B) \d E$, where $N$ is fixed. This equation needs to be solved for each $B$. The oscillating part of the density of states $D_\text{osc}$ then leads to an oscillating chemical potential $\mu + \mu_\text{osc}$ where $\mu_\text{osc} \propto A_1 \sin(2\pi F_1/B) R_{\mathrm{T}}(m_1) + A_2 \sin(2\pi F_2/B) R_{\mathrm{T}}(m_2)$. In 3D the amplitudes scale as $A_i \propto B^{3/2}$. Similar to MI terms of the form 
\begin{align}
    \cos \left(2\pi\frac{F_i(\mu+\mu_\text{osc})}{B} \right) \approx& 
    \cos \left(2\pi\frac{F_i}{B}\right) - \sin \left(2\pi\frac{F_i}{B}\right) \nonumber \\ &\times \frac{2\pi m_i}{\hbar e B} \mu_\text{osc}
\end{align}
lead to combination frequencies $F_1\pm F_2$ which again vanish fast with increasing temperature as $R_{\mathrm{T}}(m_1)R_{\mathrm{T}}(m_2)$. As a function of magnetic field, CPOs scale with the product of the Dingle factors to first order.

{\cblue Sizeable CPOs require a large fraction of the total DOS to change occupation, when the Landau tubes pass through the Fermi energy, typically realized in quasi-2D and 2D systems.}
In 3D metals however, CPOs are strongly suppressed (the amplitudes $A_i$ go with a positive power of $B$) due to the DOS tails of the Landau tubes arising from the $k$-dispersion along the field direction and the joint effect of multiple FS sheets, which act as a reservoir.

\subsection{Quantum interference (QI)}
{\cblue In this section, it is necessary to distinguish a semiclassical {\it path}, which is a 1D line segment of a FS associated with a charge carrier travel direction determined by the direction of the magnetic field and the Fermi velocity, and a semiclassical {\it orbit}, i.e., a closed semiclassical path. Quantization of semiclassical orbits gives rise to quantum oscillations, and interference between (open) paths leads to}
quantum interference (QI) oscillations, also known as Stark oscillations \cite{1971_Stark_PhysRevLett}. They arise from coherent {\cblue semiclassical paths} \cite{1984_Bergmann} between {\cblue semitransparent} MB junctions, effectively enclosing flux in an interferometer setup \cite{1983_Kaganov_PhysicsReports}. The observation of QI requires a FS which (i) realizes an interferometer setup, i.e., at least two semiclassical paths on which particles travel {\cblue between MB junctions} and which enclose an area $\mathcal{S}$ in momentum space; (ii) MB junctions between the semiclassical paths with intermediate MB probabilities $p$ leading to a partial MB with $p(1-p) \neq 0$.

We show a corresponding minimal model, comprising a 1D chain of circular orbits coupled by a weak periodic potential in Fig.\,\ref{fig:coventionalMech}(d) \cite{Deutschmann2001}. {\cblue The enclosed area of the interferometer $\mathcal{S}$ (blue shaded area in Fig.\,\ref{fig:coventionalMech}(d)) does not correspond to a closed orbit, as indicated by the red arrows along path a and b on the boundary of $\mathcal{S}$. The QI oscillations thus do not arise from Landau quantization but the interference between the two pathways a and b: Quasiparticles starting at the upper MB junction (yellow circle) follow either path a or path b, depending on whether a tunneling event occurs or not. They rejoin at the lower MB junction and their coherent superposition may lead to constructive or destructive interference, depending on the relative phases they acquired.}

In real space, the projections of the quasiparticle trajectories are rotated by $\pi/2$ and scaled with $\ell_B^2 = \frac{\hbar}{eB}$ if compared to the reciprocal-space paths \cite{1984_Shoenberg_Book}. Evaluating the transmission probability from one MB junction to the other, the quasiparticles {\cblue may take path a or path b}, picking up a relative phase $e B \mathcal{S} \ell_B^4 = \mathcal{S}/eB$. The phase can be attributed to the Aharanov--Bohm effect where the enclosed area $\mathcal{S} \ell_B^4$ depends on the magnetic field. This magnetic field dependence becomes visible through QI oscillations being periodic in $1/B$ with a frequency $\mathcal{S}$ in transport quantities. QI oscillations manifest themselves in transport properties like resistivity and not in thermodynamic properties.

QI oscillations do not naturally occur at difference frequencies of conventional QO frequencies. However, in the minimal model of Fig.\,\ref{fig:coventionalMech}(d), QOs at $F_2-F_1$ (gray area) arise when one defines $F_1$ as the frequency of the lentil-shaped orbit and $F_2$ as the frequency corresponding to the circular area. Note that $F_2$ does not correspond to a fundamental Onsager orbit, as it already requires MB to occur.

The thermal damping of QI oscillations is determined by the energy dependence of their frequency $\mathcal{S}(\epsilon_F)$, similar to standard QOs, see Fig.\,\ref{fig:schematicIdea}(e). QI pathways with the same direction of QP motion tend to be shifted in reciprocal space by changes in the chemical potential $\mu$, while changing the mutual enclosed area $\mathcal{S}$ only to second order. Therefore, QI oscillations may exhibit a very weak thermal damping~\cite{1983_Kaganov_PhysicsReports}. In situations where QI frequencies $F_{QI}=\eta_1 F_1 + \eta_2 F_2$ ($\eta_i \in \mathbb{R}$) can be expressed via QO frequencies $F_1,F_2$, the $T$-dependence is governed by $R_{\mathrm{T}}(\eta_1 m_1 + \eta_2 m_2)$. 

The magnetic field dependence of QI oscillations is expected to be governed by the field dependence of the MB probability, with QI being absent in both the low- and the high field limit where the MB probability is either $p\approx 0$ or $p\approx 1$.

Typically, crystal symmetries require QI frequencies to be a fraction of the difference frequency. Considering, e.g., the two electron orbits in Fig.\,\ref{fig:coventionalMech}(a) QI frequencies would occur at $F_2/2-F_1/2$.

\subsection{Comparison of frequency-generating mechanisms}

The above section\,\ref{sec:known-mechanisms} covers the most important and well-known non-Onsager frequency-generating mechanisms, while section\,\ref{sec:QPLOintro} introduced the new QPLOs. Given the large body of work on QOs, this short review is not meant to be exhaustive. Further, more specialized frequency-generating mechanisms may arise, e.g., from magnetic field-induced changes in the FS.
The mechanisms discussed here may be classified according to the preconditions for their occurrence and their properties in terms of frequency, temperature dependence, and field dependence as summarized in Tab.~\ref{tab:mechanism_overview}. Note that we always assume that at least two conventional extremal orbits are present.

{\cblue Note, in practice the distinction between transport and thermodynamic quantities made in Tab.~\ref{tab:mechanism_overview} may be subtle. For example, pulsed-field magnetization or field-modulation techniques may be influenced by inductive effects, effectively mixing transport characteristics into a predominantly magnetic signal.}

\begin{sidewaystable}
    \centering
    \begin{tabular}{|l|l|l|l|l|c|c|c|}
       \hline
       \multicolumn{1}{|c|}{Abbr} & \multicolumn{1}{|c|}{Mechanism} & \multicolumn{1}{|c|}{FS preconditions} & \multicolumn{1}{|c|}{Other conditions} & \multicolumn{1}{|c|}{Oscillatory} & \multicolumn{1}{|c|}{Simplest} & \multicolumn{1}{|c|}{$T$-dependence} & \multicolumn{1}{|c|}{$B$-dependence of}                      \\
             &                       &                                  &                         & \multicolumn{1}{|c|}{quantities}& \multicolumn{1}{|c|}{combination}&      &\multicolumn{1}{|c|}{amplitude ratio} \\
             &                               &                          &                         &           &\multicolumn{1}{|c|}{frequencies}&           &                          in units of $R_\mathrm{D}$\\ \hline
                    
       MB    & magnetic breakdown e--e/h--h  & existence of MB junctions& $p>0$                   & thermodynamic/          & $F_1 + F_2$      
             & $R_T(m_1+m_2)$         & $e^{-B_0/B}$\\
             &                       &                                  &                         & transport               &                  &                       &                           \\ \hline
                    
       MB    & magnetic breakdown e-h        & existence of MB junctions& $p>0$                   & thermodynamic/          & $F_1 - F_2$      
             & $R_T(|m_1|+|m_2|)$    & $e^{-B_0/B}$\\
             &                               &                           &                         & transport              &                  &                       &                           \\ \hline
       MI    & magnetic interaction          & none                     & $\mu_0\d M/\d B\sim 1$  & thermodynamic/          & $F_1 \pm F_2$   
             & $R_T(m_1) R_T(m_2)$   & $B^{-3/2}$\\ 
            &                               &                           &                         & transport               &                  &                       &                            \\ \hline
       CPO   &chemical potential oscillations& quasi-2D FS              & close to quantum        &thermodynamic/           &$F_1 \pm F_2$  & $R_T(m_1) R_T(m_2)$   &$B^{1/2}$\\ 
             &                               &                          &   limit $F_i \sim B$    & transport               &                 
             &                       &                           \\ \hline
       QI    & Stark quantum interference    & interferometer geometry  & partial MB $p (1-p)>0$  & transport & $F_1 - F_2$                   
             & $R_T(m_1 - m_2)$      & $e^{-B_0/B}(1-e^{-B_0/B})$\\
             &                               & of quasiparticle paths   &                         &                         &                  &                       &                           \\
             &                               & between MB junctions     &                         &                         &                  &                       &                           \\ \hline
        QPLO &quasiparticle lifetime oscillation& none                 & orbits coupled by       & transport               & $F_1 \pm F_2$   
             & $R_T(m_1 \pm m_2)$    & $B^{-1/2}$\\
             &                       &                                  & scattering              &                         &                 &                       &                           \\ \hline
    \end{tabular}
    \caption{Mechanisms generating sum and difference frequency combinations that do not correspond to Onsager orbits and the necessary conditions for their occurrence. For a given MB breakdown junction, $p$ quantifies the probability of an interband tunneling event. The only mechanisms that may generate QOs with strongly suppressed temperature dephasing are QI and QPLOs. The preconditions for QI to occur are very stringent, whereas QPLO may occur generically. Note that only the most simple frequency combinations and $T$ and $B$-dependences are given.}\label{tab:mechanism_overview}
\end{sidewaystable}

\section{Materials}\label{sec:materials} 

In this section, we summarize observations of QOs reported in the literature that may represent QPLOs. Some of the frequencies have already been identified as having unknown origin, others have been associated with putative FS pockets. From the data presented in the original publications it is not necessarily possible to make definite statements about the origin of the anomalous frequencies. However, the phenomenology described in the publications has features similar to the ones observed in CoSi \cite{huber2023quantum} and as predicted theoretically \cite{leeb2023theory,huber2023quantum} for \mbox{QPLOs}. 

We list the materials in loose succession, starting out with candidates where the published experimental results are most suggestive of QPLOs. Here, the authors identify forbidden difference or sum frequencies in their data and observe strongly suppressed temperature dependencies of the QO amplitude in a transport quantity. In the materials candiates further down the list, the information presented in the publications is less clear, though still suggestive of possible QPLOs.

Our objective is to stimulate further research for the correct identification of the electronic structure. 
We hope that these observations will motivate more detailed studies in a wide range of material classes, see Tab.~\ref{tab:material_overview} for an overview. Where available, we show figures reproduced from the original publications. In the QO spectra, we mark features that may be attributed to QPLOs by an orange arrow.

\subsection{CoSi}
\label{mat:CoSi}

\begin{figure*}
    \centering
    \includegraphics[width=0.9\textwidth]{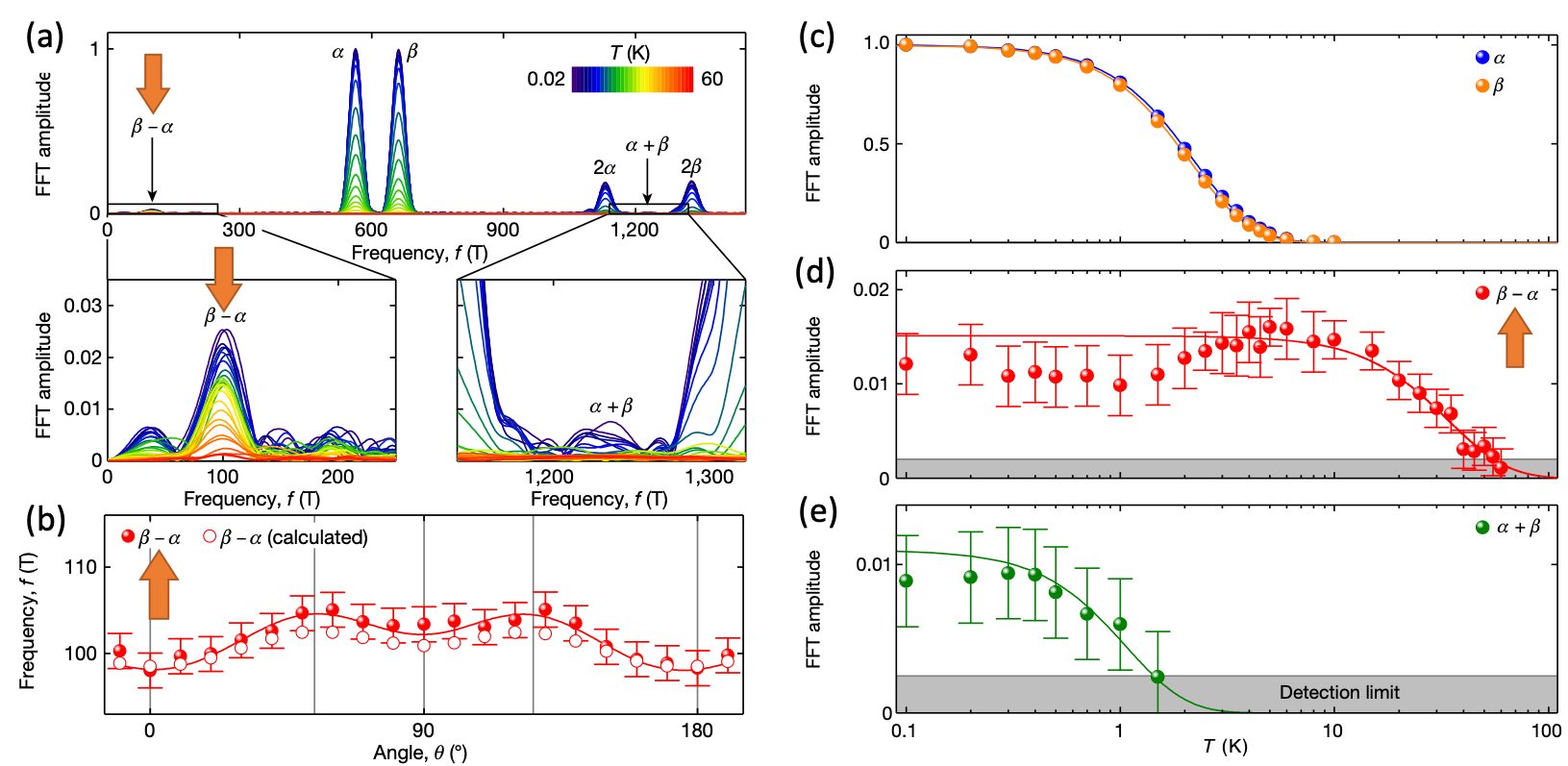}
    \caption{QPLO frequencies observed in CoSi. (a) FFT spectra and close-up views of the Shubnikov-de Haas oscillations in CoSi for different temperatures. In addition to the basis frequencies $\alpha$ and $ \beta$ and their harmonics $2\alpha, 2\beta$ peaks attributed to QPLOs occur at $\beta \pm \alpha$. (b) The measured angular dispersion of the difference frequency closely follows the difference frequency calculated from $\alpha$ and $\beta$. (c) to (e) Temperature dependence of the FFT amplitudes for $\alpha, \beta$ and $\beta \pm \alpha$. The temperature dependence of the combination frequencies $\beta \pm \alpha$ follows an LK damping term with effective masses $m_\beta \pm m_\alpha$. Reproduced with permission.\textsuperscript{\cite{huber2023quantum}} \citeyear{huber2023quantum}, Springer Nature. }
    \label{fig:CoSi}
\end{figure*}

QPLOs were discovered in the chiral topological semimetal CoSi condensing in space group 198 as published in Ref.~\cite{huber2023quantum}.

Key for the observation of QPLOs in CoSi are the extremely simple and well-understood quantum oscillatory spectra associated with the R-point of the simple cubic Brillouin zone \cite{wu2019single,xu2019crystal,wang2020haas,guo2022quasisymmetryprotected,2022_Wilde_NV,2022_Huber_PhysRevLett} comprising only two strong QO frequency branches $\alpha$ and $\beta$. QOs arising from the $\Gamma$-centered Fermi surface sheets \cite{sasmal2022shubnikovde,huber2024fermi} do not play a role in this discussion. In Ref.~\cite{huber2023quantum}, additional FFT peaks at frequencies corresponding to $\beta-\alpha$ and $\alpha+\beta$ were observed in the Shubnikov-de Haas effect, presumably because of an improved signal-to-noise ratio as compared to previous works \cite{wu2019single,guo2022quasisymmetryprotected,sasmal2022shubnikovde}, see Fig.~\ref{fig:CoSi}\,(a). A comparison of the angular dependence of the peak at $\beta \pm \alpha$ with the calculated difference of the main frequencies is shown in Fig.~\ref{fig:CoSi}\,(b) and confirms the assignment as combinations of the basis frequencies. Strikingly, the temperature dependence of the frequencies $\beta \pm \alpha$ shown Fig.~\ref{fig:CoSi}~(c)-(e) is governed by $R_T(m_\beta \pm m_\alpha)$.

For CoSi, all conventional mechanisms can be ruled out as the origin of the difference frequency as follows. The MI condition $\mu_0 \d M_\text{osc}/\d B \approx 10^{-4}$ is far below 1. Furthermore, MI oscillations exhibit a strong temperature dependence not consistent with the observation $R_T(m_\beta - m_\alpha)$. CPOs can be ruled out because the FS is strongly three-dimensional and multi-sheeted and the QOs at $\alpha$ and $\beta$ are far from the quantum limit in the maximum applied field of $\sim 18$T. Again, also the temperatue dependence of the difference frequency rules out CPOs. MB does not allow a frequency at  $\beta-\alpha$, because (i) both orbits are electron like, (ii) there is no partial magnetic breakdown with $0<p<1$ \cite{guo2022quasisymmetryprotected,2022_Wilde_NV}, but - depending on the field direction - either full MB or no MB at all.
The conditions for QI are not met, because there is full magnetic breakdown (or even no breakdown junction for specific symmetry directions).
Even when assuming partial breakdown, the QI interferometer areas between breakdown junctions do not correspond to $\beta-\alpha$.

In contrast, the experimental observation of a difference and a sum frequency $\beta-\alpha$ and $\alpha+\beta$ with temperature dephasing governed by $R_T(m_\beta - m_\alpha)$ and $R_T(m_\beta + m_\alpha)$, respectively, are consistently explained by the QPLO mechanism.

\subsection{\texorpdfstring{MoSi$_2$}{MoSi2}}
\label{mat:MoSi}
Pavlosiuk et al. \cite{pavlosiuk2022giant} report measurements of Shubnikov--de Haas oscillations in the topological semimetal MoSi$_2$. MoSi$_2$ crystallizes in the tetragonal spacegroup 139. The Fermi surface comprises a pillow-shaped electron pocket around the $Z$ point and a pill-shaped hole pocket around the $\Gamma$ point. 

\begin{figure}
    \centering
    \includegraphics[width=0.9\columnwidth]{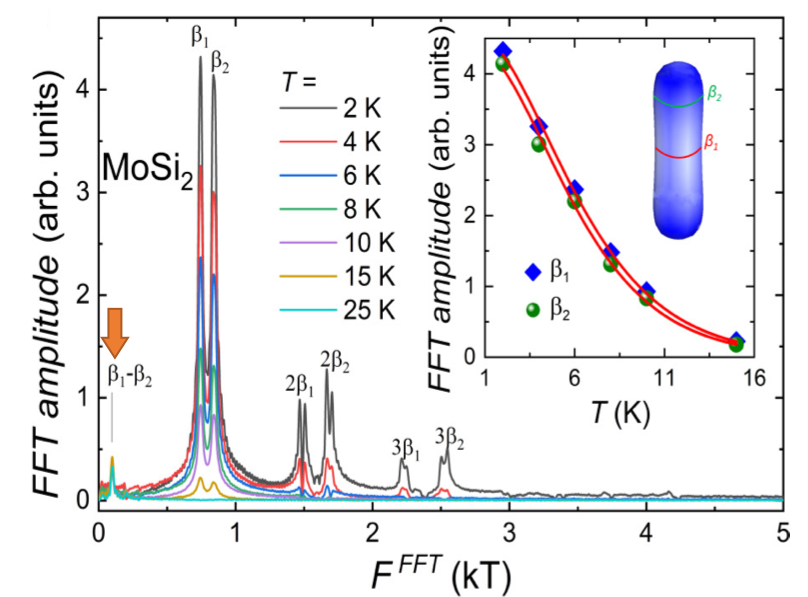}
    \caption{Spectrum of Shubnikov-de Haas oscillations of MoSi$_2$ for different temperatures and $B\parallel c$. The inset shows the orbits of the main frequencies $\beta_1$ and $\beta_2$ around the pill-shaped hole pocket. The temperature dependence of the oscillation amplitudes reveal that the two frequencies have almost identical effective masses. A difference frequency is observed up to high temperatures and becomes the dominant frequency of the spectrum above 15K. Reproduced with permission.\textsuperscript{\cite{pavlosiuk2022giant}} \citeyear{pavlosiuk2022giant}, American Physical Society. }
    \label{fig:MoSi}
\end{figure}

The Shubnikov-de Haas spectra reported in Ref.~\cite{pavlosiuk2022giant} for field along the c-axis are shown in Fig.~\ref{fig:MoSi}. They feature two strong, close-lying main frequencies $\beta_1$ and $\beta_2$ and several higher harmonics. The main frequencies were identified as the extremal orbits of the pill-shaped hole pocket as shown in the inset. The authors of Ref.~\cite{pavlosiuk2022giant} argue that the QO frequency associated with the pillow-shaped-electron pocket cannot be observed due to its large effective mass.

Additionally, Pavlosiuk et al. report the observation of a difference frequency $\beta_2-\beta_1$ up to exceptionally high temperatures. The assignment as a difference frequency is confirmed by the angular dependence. 

The authors note that this difference is unusual and tentatively attribute it to MI (called the Shoenberg effect in Ref.~\cite{pavlosiuk2022giant}). However, the authors point out that MI may only be observed at much reduced temperatures, and that the temperature dependence of the difference frequency is not consistent with MI. While the authors do not show a curve with the temperature dependence, it may be inferred from the $T$-dependence of the FFT amplitudes, shown in Fig.~\ref{fig:MoSi}. It shows that the difference frequency has a large amplitude at 25 K where the fundamental frequencies are already suppressed completely.

{\cblue Interestingly, the difference frequency oscillations are visibly absent in the dHvA spectra reported in Ref.~\cite{van1987haas} while the sum frequencies are attributed to MI.}

Comparison with the mechanism for QPLOs reveals remarkable similarities in all aspects including the inconsistencies with the conventional non-Onsager mechanisms of combination frequencies. The FS geometry does not feature any MB junctions, and orbits are located far apart from each other in the Brillouin zone such that MB and QI-induced combination frequencies can be ruled out. The main frequencies at $\sim 1$kT are very far away from the quantum limit, rendering CPOs irrelevant. The observed temperature dependence is at odds with all other mechanisms. The fact that no difference frequency of $\beta_1$ and $\beta_2$ was observed in the de Haas--van Alphen effect in MoSi$_2$ \cite{van1987haas} is in accordance with the theory of QPLOs and at odds with MB, MI and CPOs. 

Remarkably, the corresponding fundamental frequencies arise from two extremal cross sections of the same FS pocket, i.e., interorbit coupling in the QPLO mechanism is here mediated by intraband coupling and not by interband coupling.

\subsection{\texorpdfstring{WSi$_2$}{WSi2}}
\label{mat:WSi}
The electronic band structure of the topological semimetal WSi$_2$ is very similar to MoSi$_2$ \cite{pavlosiuk2022giant}. They share the same FS structure, consisting of a pill-shaped hole pocket around the $\Gamma$ point and a pillow-shaped electron pocket around the $Z$ point.

\begin{figure}
    \centering
    \includegraphics[width=0.9\columnwidth]{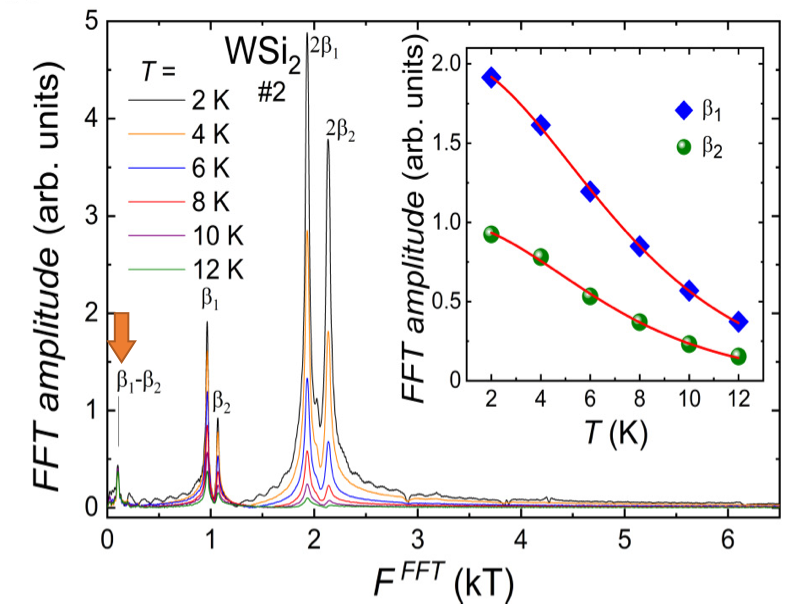}
    \caption{Spectrum of Shubnikov-de Haas oscillations of WSi$_2$ for different temperatures and $B\parallel c$. The main frequencies feature almost identical masses as shown in the inset. A difference frequency is observed up to high temperatures and becomes the dominant frequency of the spectrum above 10K. Reproduced with permission.\textsuperscript{\cite{pavlosiuk2022giant}} \citeyear{pavlosiuk2022giant}, American Physical Society.}
    \label{fig:WSi}
\end{figure}

The Shubnikov-de Haas spectrum for $B\parallel c$ reported in Ref.~\cite{pavlosiuk2022giant} is shown in Fig.~\ref{fig:WSi}. It resembles the one of MoSi$_2$. The assignment of orbits is analogous, i.e.,  $\beta_1$ and $\beta_2$ are identified as the two extremal orbits around the pill-shaped hole pocket. In WSi$_2$, the dominating QO frequencies are the 2nd harmonics and not the 1st harmonics. This might be suggestive of strong scattering but is more likely related to a large Zeeman effect \cite{pavlosiuk2022giant}. The authors report the observation of a temperature-stable difference frequency $\beta_1-\beta_2$, see Fig.~\ref{fig:WSi}. The frequency can already be seen in the bare resistivity data at 25K. Additionally, there is a weak peak visible in the spectrum between the second harmonics which could be related to the sum $\beta_1+\beta_2$.{\cblue No difference frequency is visible in the dHvA spectra reported in Ref.~\cite{mondal2020WSi2}.}

Following the same arguments as for MoSi$_2$ above, MB, MI, CPOs, and QI may be ruled out as a source of the difference frequency. In contrast, QPLOs consistently explain the reported observations including the suppressed temperature dephasing. Again, interorbit coupling is here mediated by intraband coupling, because the corresponding fundamental frequencies arise from two extremal cross-sections of the same Fermi surface pocket.

\subsection{\texorpdfstring{Co$_3$Sn$_2$S$_2$}{Co3Sn2S2}}
\label{mat:CoSnS}
	
The ferromagnetic kagome semimetal Co$_3$Sn$_2$S$_2$, condensing in the rhombohedral spacegroup 166, features twelve banana-shaped electron pockets (E1) across the Brillouin zone, six tetrahedron-like electron pockets (E2), six cylinder-like hole pockets (H1) across the Brillouin zone and six tetrahedron-like hole pockets (H2) \cite{ding2021quantum}. 

\begin{figure}
    \centering
    \includegraphics[width=\columnwidth]{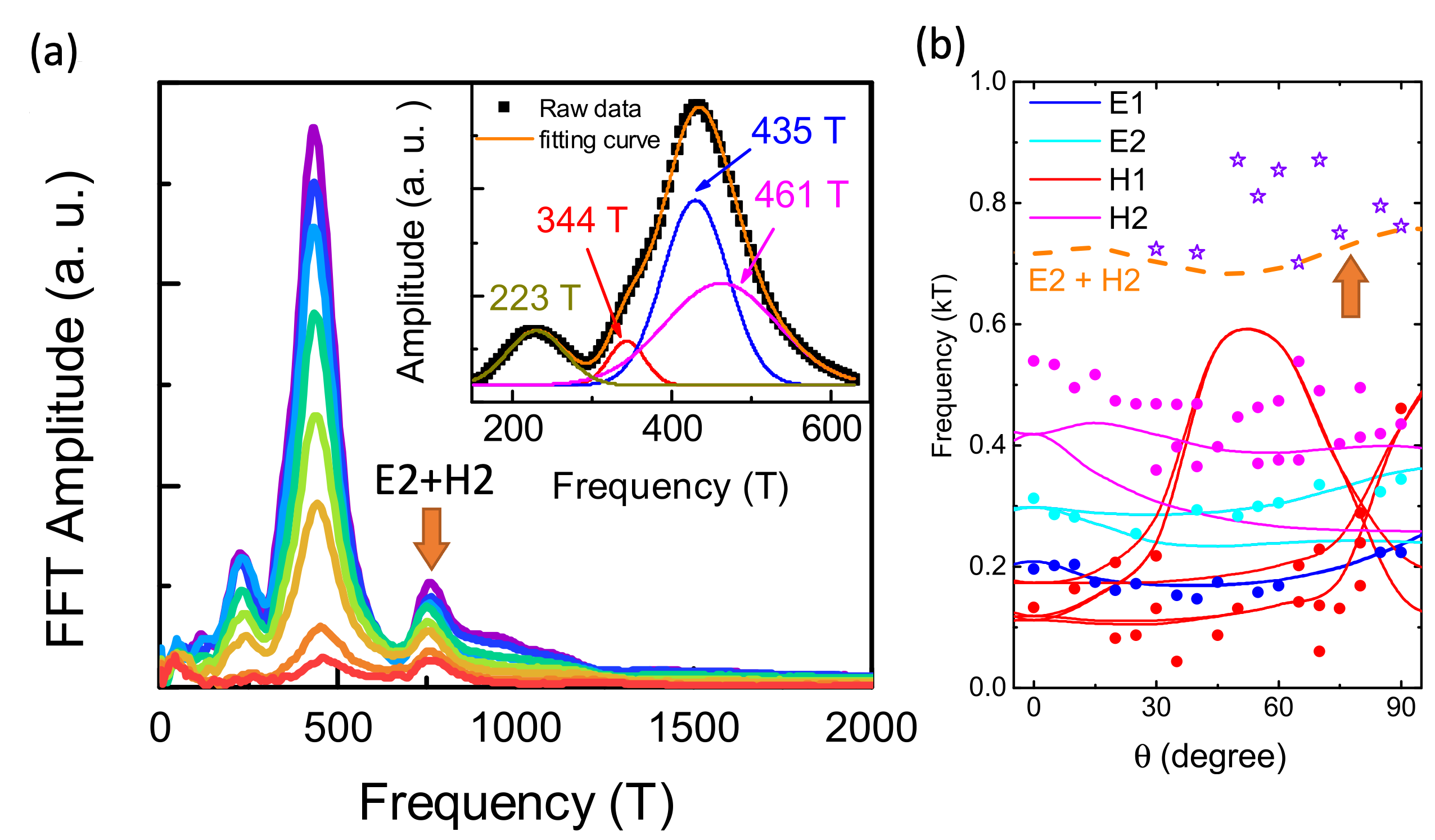}
    \caption{(a) SdH spectrum of the ferromagnetic semimetal Co$_3$Sn$_2$S$_2$ for various temperatures from 1.6 K (purple) to 20.1 K (red). (b) Angular dependence of the peak frequencies. Reproduced with permission.\textsuperscript{\cite{ding2021quantum}} \citeyear{ding2021quantum}, IOP Publishing Ltd. }
    \label{fig:CoSnS}
\end{figure}

The SdH spectrum of the resistivity for field direction $B~||~y$-axis is shown in Fig.~\ref{fig:CoSnS}~(a). The authors analyzed the individual frequencies via a Gauss multi-peak fit, yielding $\mathrm{H1} = 214 \mathrm{T}$, $\mathrm{E2} = 343 \mathrm{T}$ and $\mathrm{H2} = 429 \mathrm{T}$. An additional peak at 760 T, which decreases slowly with increasing temperature, is clearly visible. The 760 T peak is identified by the authors as a sum frequency $\mathrm{E2}+\mathrm{H2}$ of an electron and a hole orbit. 

Considering the presence of electron and hole Fermi surface sheets, the authors explain the sum frequency semiclassically by Klein tunneling and a partial magnetic breakdown, even though this would result in a strongly temperature-dependent difference frequency $E2-H2$ \cite{obrien2016magnetic} and not in a temperature-stable sum frequency. In fact, the sum frequency observed in Ref.\,\cite{ding2021quantum} is semiclassically forbidden because the quasiparticle orbit would oppose the Lorentz force on a part of the trajectory. Additionally, the spectrum shows that at 20 K the sum frequency becomes the dominant part of the QO spectrum, i.e., its temperature dependence is markedly weaker than that of E2 and H2. 

Nonetheless, the authors attribute the sum frequency to MB. Yet, they explicitly mention that magnetic breakdown can only explain a difference frequency. For an alternative explanation, MI and CPOs can be ruled out, due to the weak temperature dependence. The behavior of the sum frequency may instead be explained consistently in terms of QPLOs, predicting $m_{\mathrm{E}2+\mathrm{H}2}=0.11 m_e$. This would then represent interband coupling of electron and hole orbits as the cause of a sum frequency with light mass, in contrast to two electron orbits leading to a light difference frequency. It is unclear whether QI is a possible alternative explanation which would require an analysis of the MB junctions.

\subsection{HfSiS}
\label{mat:HfSiS}
\begin{figure*}
    \centering
    \includegraphics[width=0.9\textwidth]{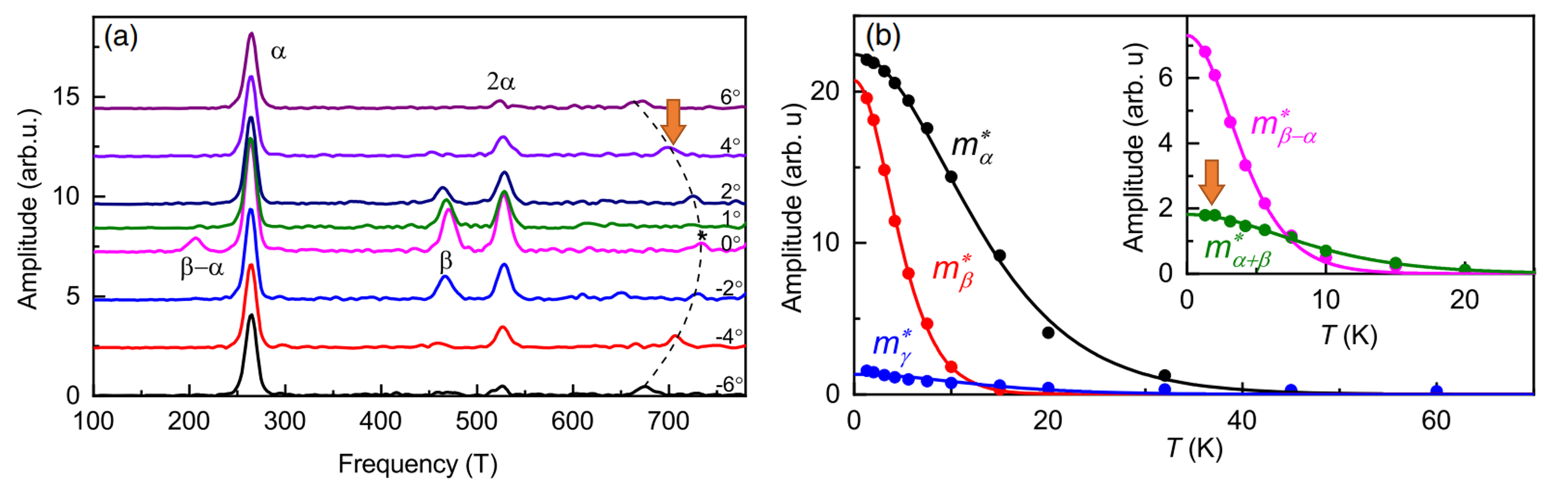}
    \caption{(a) SdH spectra of the nodal-line semimetal HfSiS. The sum frequency $\alpha + \beta$ is marked by the dashed line. It is observed in a much larger angular range than the difference frequency which is due to electron-hole MB. (b) Temperature dependence of the oscillation amplitudes yielding cyclotron masses of $m_{\alpha}=0.177m_e$, $m_{\beta}=0.48m_e$ and $m_{\alpha+\beta}=0.22m_e$. Reproduced with permission.\textsuperscript{\cite{vandelft2018electronhole}} \citeyear{vandelft2018electronhole}, American Physical Society.}
    \label{fig:HfSiS}
\end{figure*}

The quantum oscillation spectra in the tetragonal (SG 129) nodal-line semimetal HfSiS, reported in Ref.\,\cite{vandelft2018electronhole}, are dominated by extremal orbits of a hole (frequency $\alpha)$ and an electron (frequency $\beta$) FS sheet, see Fig.~\ref{fig:HfSiS}~(a). Additional combination frequencies at the difference, $\beta-\alpha$, and the sum, $\alpha+\beta$, are observed in the SdH spectrum of the longitudinal resistance. Whereas the difference frequency $\beta-\alpha$ can be explained as MB between the hole and the electron orbit, the authors point out explicitly that the sum frequency they observe is semiclassically forbidden. Thus, MB can be ruled out. Furthermore, the sum frequency exhibits an effective mass roughly corresponding to the difference of $m_\alpha$ and $m_\beta$.

{\cblue A later study by the same group probing the dHvA effect via torque magnetometry reports the detection of MB frequencies corresponding to $n\beta - p\alpha$ \cite{mueller2022HfSiS}. However, no sum frequency $\alpha + \beta$ is detected in the dHvA spectra.}

In Ref.\,\cite{vandelft2018electronhole} the authors tentatively attribute the peak at $\alpha+\beta$ to MI, but note that the magnetization they determined for their samples is far too low to account for the combination frequencies they observe. In addition, the temperature dephasing is too weak for this interpretation. CPOs can be ruled out because of the 3D multi-sheeted FS and QOs far away from the quantum limit. QI is very unlikely, as it requires partial MB. However, the authors show in detail that the difference frequency (which is due to MB) may be observed only in a very narrow angular range around the c-axis. Their calculation of the breakdown probability from DFT explains this behavior, showing that MB is already suppressed by an order of magnitude for $~1^{\circ}$ misalignment with the c-axis. In contrast, the sum frequency is observed in a much wider angular range and shows a weaker temperature dependence. Importantly, the phenomenology of the frequency at $\alpha+\beta$ is precisely what is expected for QPLOs between an electron and a hole orbit, i.e., an electron-hole sum frequency with a temperature dependence governed by the difference of the masses. 

\subsection{ZrSiS}
\label{mat:ZrSiS}
\begin{figure}
    \centering
    \includegraphics[width=\columnwidth]{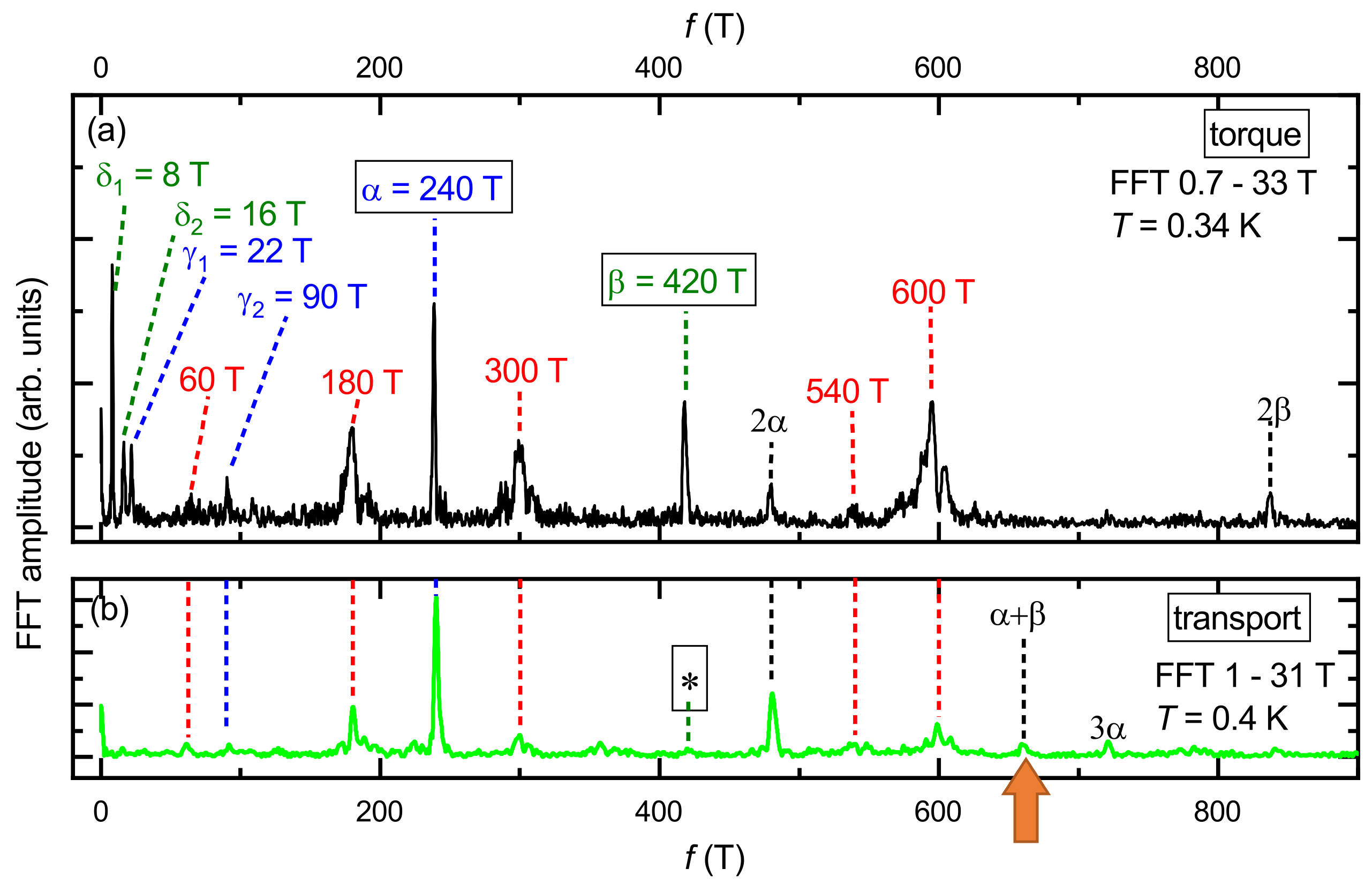}
    \caption{Quantum oscillatory spectra of the 3D topological semimetal ZrSiS. (a) de Haas--van Alphen spectra (black). (b) Shubnikov--de Haas effect (light green). Fundamental frequencies due to hole sheets are labeled in blue. Fundamental frequencies due to electron sheets are labeled in green. Higher harmonics are labeled in black. Frequencies attributed to MB are labeled in red. A semiclassically forbidden sum frequency $\alpha+\beta$ is observed only in the transport data, consistent with an origin from QPLOs. Reproduced with permission.\textsuperscript{\cite{muller2020determination}} \citeyear{muller2020determination}, American Physical Society. }
    \label{fig:ZrSiS}
\end{figure}

Belonging to the same material family as HfSiS, the semimetal ZrSiS also features a Fermi surface with hole pockets ($\alpha$) and electron pockets ($\beta$). Multiple breakdown frequencies (marked in red) are reported in Ref.\,\cite{muller2020determination} in the torque magnetization (de Haas--van Alphen effect) and transport quantities (Shubnikov-de Haas effect), see Fig.~\ref{fig:ZrSiS}. Additionally, a sum frequency $\alpha+\beta$ is observed only in the Shubnikov-de Haas effect and not in the dHvA effect. Analogous to the case of HfSiS above, the sum frequency cannot be explained by MB between electron and hole orbits, which would yield only a difference frequency. No data on the angular dispersion and temperature dependence of the sum frequency in the SdH effect is reported. The sum frequency may be explained in terms of QPLOs analogous to HfSiS. The fact that this semiclassically forbidden frequency is observed only in the SdH effect and not in the dHvA effect is a hallmark of QPLOs, in accordance with, both, the analytical theory \cite{leeb2023theory,huber2023quantum} and numerical simulations \cite{leeb2024interband, leeb2024numerical}.
Here, it is unclear whether QI is a possible alternative explanation which would require an analysis of the MB junctions. However, it may be possible to rule out QI experimentally by comparing the angular dependence of the sum and the difference frequency as discussed above for HfSiS.

{\cblue
\subsection{\texorpdfstring{KV$_2$Se$_2$O}{KV2Se2O}}
The oxyselenide KV$_2$Se$_2$O belongs to a class of layered 3$d$ transition-metal compounds forged to improve the understanding of high-temperature superconductivity in structurally related cuprates \cite{bai2024absence}. The compound caused significant attention recently because it has been proposed to host altermagnetism at room temperature \cite{jiang2024discovery}.

Ref.~\cite{bai2024absence} reports on SdH oscillations in the magnetoresistance which can be discerned up to 30 K. The authors identify 4 main frequencies $\eta, \beta,\gamma, \delta$ based on the height of the FFT peaks. They could not assign the frequencies to the pockets obtained from first-principle DFT calculations of non-magnetic KV$_2$Se$_2$O. Additionally, they identify peaks $\gamma-\beta, \eta-\delta, \beta+\eta, \gamma + \eta$ which they associate with combinations of the main frequencies, see Fig~\ref{fig:KVSeO}. The authors associate the combination frequencies to MB and QPLOs without further discussion. 

From the FFTs shown in Fig.~\ref{fig:KVSeO}(a), it can be seen that the combination frequencies $\eta-\delta$ and $\beta+\eta$ exhibit a smaller effective mass than the heavy $\eta$ orbit, indicating that they cannot be explained by MB, MI, or CPOs but pointing towards QI or QPLOs. Additionally, CPOs can be ruled out because at least one of the contributing frequencies is of the order of a few hundred Tesla, which is significantly larger than the maximal field of $16$ T, i.e., it is far from the quantum limit. Without a calculated FS allowing a correct assignment of the QO orbits, it is not possible to assign specific combination frequencies to MB, QI, and QPLOs. However, we note that MB alone cannot explain all combination frequencies, no matter which orbits are electron- or hole-like. Future studies will need to show what the origin of combination frequencies in KV$_2$Se$_2$O is.

\begin{figure}
    \centering
    \includegraphics[width=\columnwidth]{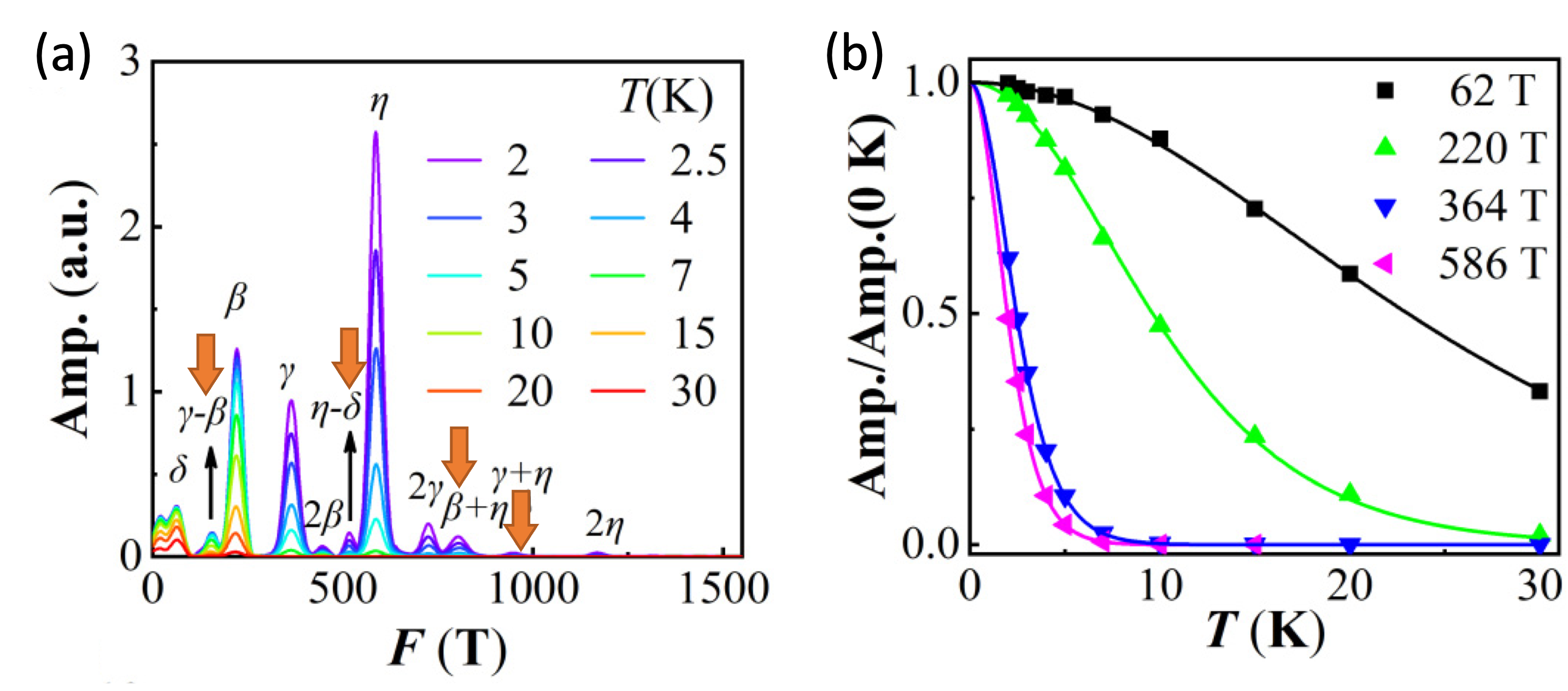}
    \caption{{\cblue Shubnikov-de~Haas oscillations in KV$_2$Se$_2$O. (a)~FFT spectra with four fundamental frequencies and several combinations thereof. (b)~Amplitudes of the fundamental frequencies as a function of temperature. No analysis of the temperature dependence of the combination frequencies is given. Reproduced with permission.\textsuperscript{\cite{bai2024absence}} \citeyear{bai2024absence}, Springer Nature.}}
    \label{fig:KVSeO}
\end{figure}
}

\subsection{\texorpdfstring{FeSe$_{1-x}$S$_x$}{FeSe(1-x)S(x)}}
\begin{figure}
    \centering
    \includegraphics[width=\linewidth]{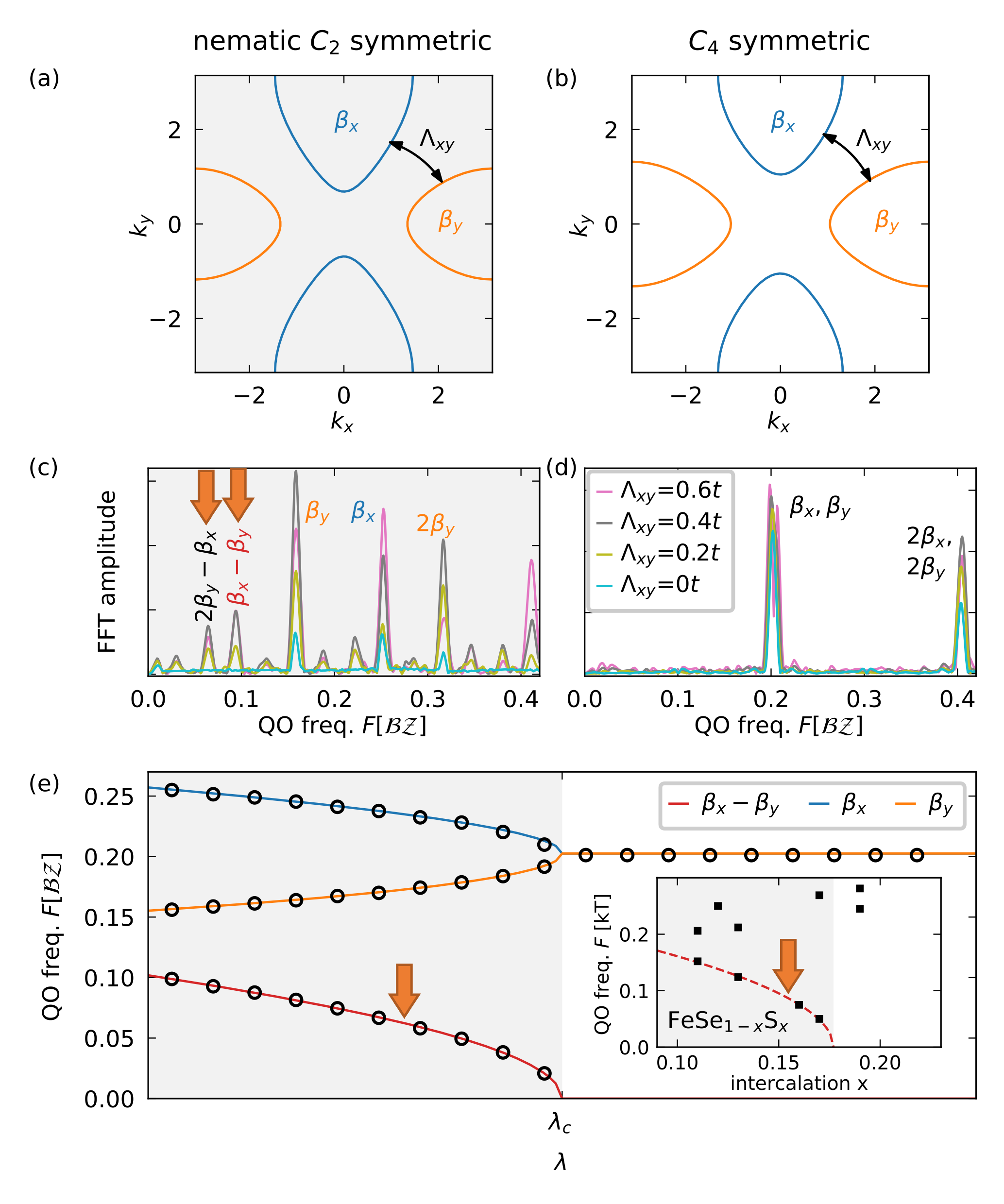}
    \caption{
    Schematic depiction of the FS of FeSe$_{1-x}$S$_x$ in the symmetric, tetragonal phase (b) and in the nematic phase (a) where the $\beta_x$ and the $\beta_y$ pockets have different size. The pockets are assumed to be coupled by interband coupling parameterized by $\Lambda_{xy}$. The numerically calculated SdH QO spectra for the nematic FS (c) and the symmetric, tetragonal phase (d). (c) shows a splitting of the main frequency, as semiclassically expected, and additionally combination frequencies which can be attributed to QPLOs. Extracting the QO frequency when moving towards the phase transition (e) shows similar behavior as measured in FeSe$_{1-x}$S$_x$ (inset) \cite{coldea2019evolution}.
    Reproduced with permission.\textsuperscript{\cite{leeb2024interband}} \citeyear{leeb2024interband}, American Physical Society. }
    \label{fig:FeSe}
\end{figure}

The iron-based superconductor FeSe exhibits a phase transition from a tetragonal to an electronic nematic phase below a critical temperature which decreases as a function of pressure or doping. SdH measurements in FeSe$_{1-x}$S$_x$ show an unexpected, slow frequency $\lambda$ in the nematic phase \cite{coldea2019evolution,reiss2020quenched}, see inset of Fig.\ref{fig:FeSe}~(e). The appearance of this new frequency was argued to be the hallmark of a Lifshitz transition, i.e., the formation of a new Fermi pocket. However, the proposal of the new Fermi pocket is exclusively based on the QOs observed experimentally. It is not supported by ab-initio calculations of the band structure.

Interband scattering may provide an alternative explanation where the new frequency $\lambda$ would not be related to a Lifshitz transition, as detailed in Ref.\,\cite{leeb2024interband} and briefly summarized in the following. At the nematic transition, the Fermi surface loses its $C_4$ rotational symmetry. It becomes energetically favorable for the system to acquire a spontaneous orbital asymmetry. For instance, the size of the $\beta_x$ pocket may increase whereas the size of the $\beta_y$ pocket decreases, as sketched in Fig.\,\ref{fig:FeSe}~(a) and (b). 

The QO spectrum is sensitive to such a symmetry breaking. In the tetragonal phase, the frequencies of the $\beta_x$ and $\beta_y$ pocket are essentially degenerate as shown in Fig.\ref{fig:FeSe}~(b) and (d). In comparison, the frequencies get split into two adjacent frequencies in the nematic phase, see Fig.\ref{fig:FeSe}~(a) and (c). In the presence of non-linear interband coupling a frequency $\lambda = |\beta_x-\beta_y|$ emerges in the nematic phase representing the difference of $\beta_x$ and $\beta_y$, see Fig.\ref{fig:FeSe}~(e). Thus, there is no need to assume a Lifshitz transition to account for the emergence of the additional frequency $\lambda$, which would be due to QPLOs. Clarifying the origin of the frequency $\lambda$ would provide important input on the nature of the superconducting phase in FeSe$_{1-x}$S$_x$ as discussed further in Ref.\,\cite{leeb2024interband}.

\subsection{\texorpdfstring{WTe$_2$}{WTe2}}
\begin{figure}
    \centering
    \includegraphics[width=\linewidth]{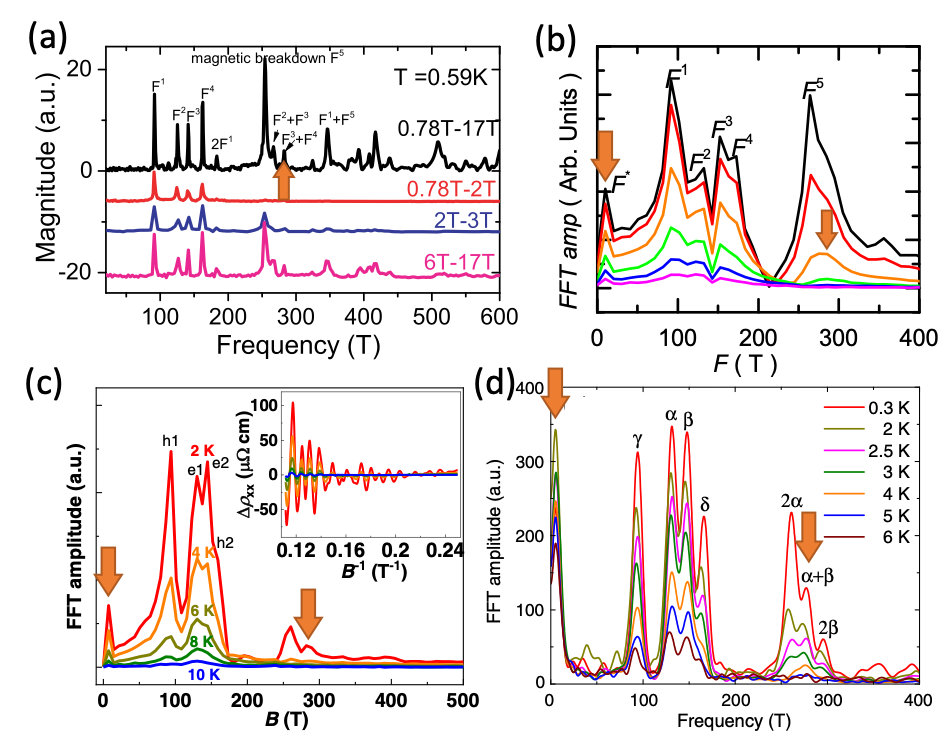}
    \caption{SdH spectra of WTe$_2$ for different temperatures. 
    (a) The FFT spectra of the Seebeck coefficient for different magnetic field windows. Reproduced with permission.\textsuperscript{\cite{zhu2015quantum}} \citeyear{zhu2015quantum}, American Physical Society.
    (b) Magnetoresistance spectra for temperatures from 1.8 K to 10 K. F$^5$ is identified as sum frequency of F$^1 + \mathrm{F}^4$. A low-frequency peak with F$^* = 10$\,T is identified with an effective mass $m_{\mathrm{F}^*} = 0.29 m_e$.  Reproduced with permission.\textsuperscript{\cite{wu2017threedimensionality}} \citeyear{wu2017threedimensionality}, American Physical Society.
    (c) Magnetoresistance spectra for different temperatures. No combination frequencies were identified by the authors but additional peaks are visible in the spectrum. Reproduced with permission.\textsuperscript{\cite{pan2022ultrahigh}} \citeyear{pan2022ultrahigh}, Springer Nature.
    (d) FFT spectra of the magnetoresistance oscillations for different temperatures. The temperature-stable low-frequency peak dominates the spectrum at $T=6$\,K. Contrary to the other works, the authors of Ref.\,\cite{cai2015drastic} claim that $\alpha$ and $\beta$ correspond to a pair of particle-hole pockets. Hence, $\alpha+\beta$ would be semiclassically forbidden. Reproduced with permission.  \textsuperscript{\cite{cai2015drastic}} \citeyear{cai2015drastic}, American Physical Society.
    }
    \label{fig:WTe}
\end{figure}

References~\cite{zhu2015quantum,wu2017threedimensionality,pan2022ultrahigh,cai2015drastic} report on quantum oscillations in the transport properties of WTe$_2$. In general, four main oscillation frequencies, F$^1$-F$^4$, are attributed to nested hole (F$^1$ and F$^4$) and electron (F$^2$ and F$^3$) pockets in Refs.~\cite{zhu2015quantum,wu2017threedimensionality,pan2022ultrahigh} as shown in Fig.~\ref{fig:WTe}. Ref.~\cite{zhu2015quantum} resolves, in addition, several combination frequencies and interprets all of them as due to MB; see Fig.~\ref{fig:WTe}~(a). While most frequencies are consistent with an MB scenario, the frequency component observed at F$^3$+F$^4$ is a combination of a particle and a hole orbit, and hence semiclassically forbidden, raising the question of another mechanism explaining its existence. In Ref.\,\cite{wu2017threedimensionality} [Fig.~\ref{fig:WTe}~(b)] and Ref.\,\cite{pan2022ultrahigh} [Fig.~\ref{fig:WTe}~(c)] peak-like features are visible at the same frequency F$^3$+F$^4$.

The assignment of orbits in Ref.~\cite{cai2015drastic} [Fig.~\ref{fig:WTe}~(d)], is at odds with the other works. The frequencies $\alpha$ ($\gamma$) and $\beta$ ($\delta$) are identified as a pair of a particle and a hole pocket. However, the authors of Ref.~\cite{cai2015drastic} also identify a peak corresponding to the sum $\alpha+\beta$ and attribute it to MB, which is semiclassically forbidden. This peak is also clearly resolved in the other works, see Fig.\ref{fig:WTe}~(a)-(c) \cite{zhu2015quantum,wu2017threedimensionality,pan2022ultrahigh}, however there it would be an allowed MB frequency, due to the different orbit assignment. 

Ref.~\cite{wu2017threedimensionality} additionally identifies a low-frequency peak with a low effective mass $0.29 m_e$. The resolution of the spectrum is insufficient to assign it unambiguously as the difference of two orbits. The relatively temperature stable low-frequency peak is also observed in Ref.~\cite{pan2022ultrahigh} and Ref.~\cite{cai2015drastic}, see Fig.~\ref{fig:WTe}~(b) and (c). {\cblue The relatively large amplitude of the low-frequency peak in (d) may be associated with the superposition with a residual non-oscillatory background, as indicated by the finite value of the spectra at zero frequency.}

WTe$_2$ features at least two frequencies, potentially related to QPLOs, a temperature-stable low-frequency peak, and a sum frequency of a particle and a hole orbit. We hope that this may motivate further studies where the data are analyzed with respect to QPLO and QI-based scenarios.

\subsection{PrAlSi}
\begin{figure}
    \centering
    \includegraphics[width=\linewidth]{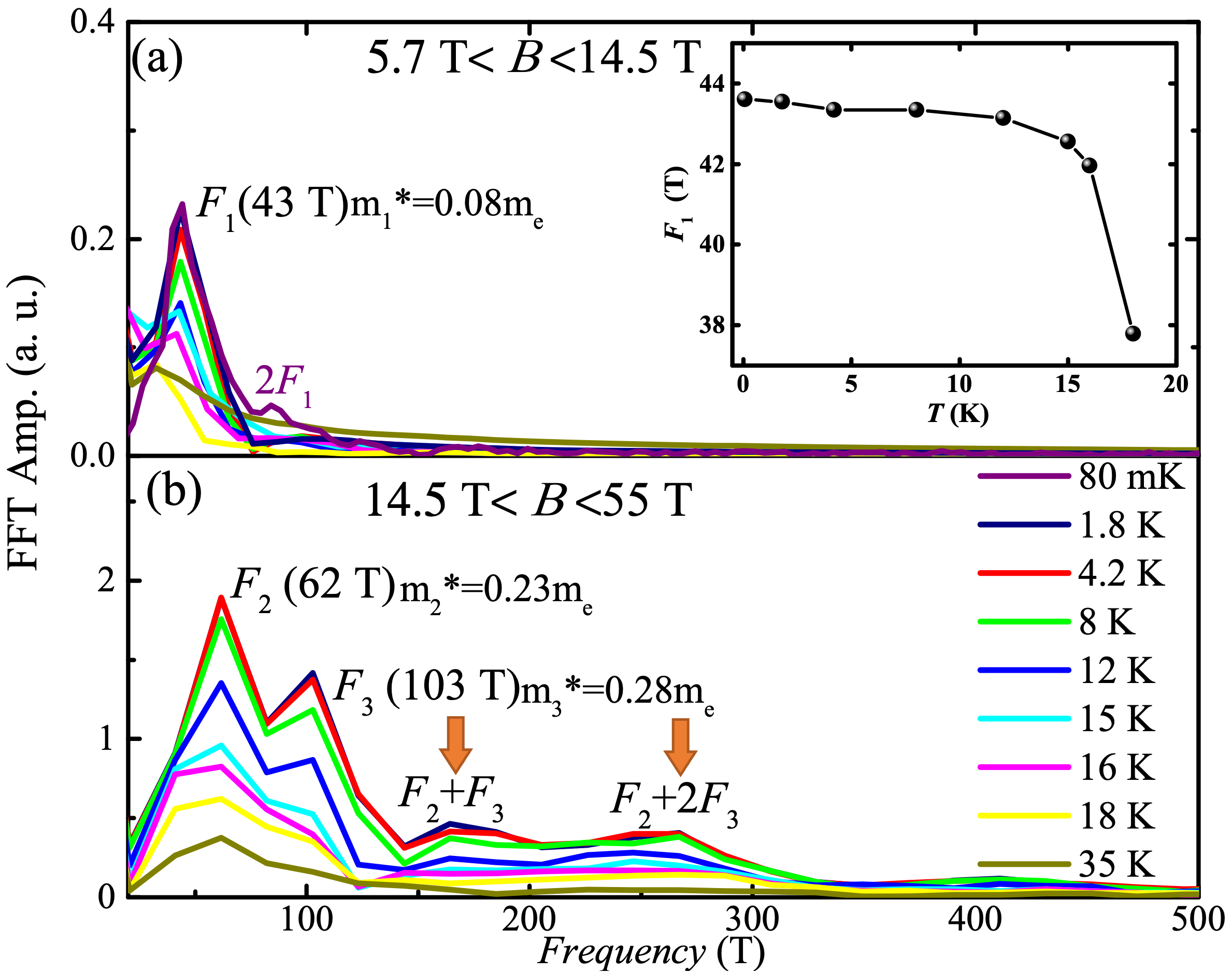}
    \caption{(a) Low-field SdH spectrum of PrAlSi where only a single QO frequency $F_1$ is observed. (b) In the high field SdH spectrum two basis frequencies are visible and associated with an electron and a hole orbit by the authors of Ref.\,\cite{wu2023fieldinduced}. The sum combinations $F_2+F_3$ and $F_2+2F_3$ are semiclassically forbidden. Reproduced with permission.\textsuperscript{\cite{wu2023fieldinduced}} \citeyear{wu2023fieldinduced}, Springer Nature.}
    \label{fig:PrAlSi}
\end{figure}

According to Ref.\,\cite{wu2023fieldinduced}, the magnetic Weyl semimetal candidate PrAlSi features 8 hole pockets which are related by symmetry such that only a single QO frequency $F_1$ is observed in the low-field regime below 14.5 T as shown in Fig.~\ref{fig:PrAlSi}~(a). Wu et al. \cite{wu2023fieldinduced} suggest that at 14.5\,T a Lifshitz transition occurs where the Weyl cones merge pairwise and form particle and hole pockets. In total, there are 4 particle and 4 hole pockets, which are again related by symmetry. They are attributed to the two main peaks $F_2$, $F_3$ in the high field ($B>14.5$ T) SdH spectrum, see Fig.~\ref{fig:PrAlSi}~(b). Additionally, there are peaks visible, which are identified as $F_2+F_3$ and $F_2+2F_3$ by the authors. However, $F_2+F_3$ and $F_2+2F_3$ are semiclassically forbidden in the scenario proposed by the authors, because the combined electron-hole orbit may only give rise to a difference frequency. Unfortunately, the published data does not allow us to infer their temperature dependence. Since the authors do not give any further information, MI, CPOs or QPLO constitute a possible explanation. The fact that $F_2+F_3$ and $F_2+2F_3$ do at least not appear to be strongly temperature dependent might be an indication against MI and CPOs. How a putative interband coupling would affect the suggested Lifshitz transition scenario is unclear. These findings may motivate more detailed studies.

\subsection{\texorpdfstring{URu$_2$Si$_2$}{URu2Si2}}
Hassinger et al. \cite{hassinger2010similarity} report two previously unobserved QO branches in the SdH spectra of the heavy-fermion superconductor URu$_2$Si$_2$ measured in the normal state of the hidden-order phase. The authors state that they observe several combination frequencies that they attribute to QI without detailed discussion. Because QPLOs may produce apparent frequency branches with similar characteristics, they may appear as an alternative explanation that is less sensitive to the FS details if compared to QI, which has stringent requirements in terms of semi-transparent MB junctions. The absence of more detailed data precludes a clear statement.

\subsection{Organic conductors}
Audouard and Fortin \cite{audouard2013organic} review the QO physics observed in layered organic conductors in high magnetic fields. The authors report in detail on combination frequencies with suppressed temperature dephasing. Most of them can be nicely explained by QI. However, the authors explicitly state that several experimental observations of combination frequencies, including SdH oscillations observable up to 32\,K which, to the authors' knowledge, constitutes a world record for an organic metal, remain unexplained. The authors claim that these combination frequencies are inconsistent with the MB-enabled network of open orbits which serves to explain all other main observations. It follows that layered organic conductors may represent example systems where QI and QPLOs are present simultaneously. A precise and careful semiclassical orbit and MB junction analysis would be needed to show which frequencies may be attributed to which effect.

    
\subsection{\texorpdfstring{UTe$_2$}{UTe2}}
Broyles et al. \cite{broyles2023revealing} report tunneling diode oscillator measurements on the spin-triplet superconductor $\mathrm{UTe}_2$. The authors observe sum and difference frequencies of electron and hole orbits and carefully discuss all potential mechanisms, apart from QPLOs, which where not known at the time. They identify MB due to Klein tunneling as origin for the difference frequency and QI as origin of the sum frequency.{\cblue Ref.~\cite{2024weinbergerUTe2} reports on the observation of further weakly temperature-dependent frequency contributions in the conductivity of $\mathrm{UTe}_2$ which are absent in the dHvA effect. The authors again explain their findings by QI.}
The QI scenario is consistent with the weak temperature dependence of the sum frequency observed experimentally. While the analysis in Refs.~\cite{broyles2023revealing, 2024weinbergerUTe2} is sound, QPLOs may represent an alternative explanation consistent with the experiment. A possible route allowing the distinction of the mechanisms may be a careful analysis of the angle-dependent breakdown probabilities, which would have a decisive influence on QI but not on QPLOs.
 

\subsection{\texorpdfstring{GdRu$_2$Si$_2$}{GdRu2Si2}}
Matsuyama et al. \cite{matsuyama2023quantum} performed magnetic torque (dHvA) and resistivity (SdH) measurements in the centrosymmetric skyrmion host GdRu$_2$Si$_2$. The authors identify 3 basis frequencies ($F_1, F_2, F_3$) in the field-polarized phase which are observed in both the dHvA effect and the SdH effect. However, the SdH spectrum is dominated by an additional slow oscillation whose spectral peak is about an order of magnitude larger. The frequency of this additional slow oscillation frequency roughly corresponds to $F_2-F_1$, the effective mass approximately to $m_2-m_1$. It is observed in a very limited angular range only. The authors do not assign this frequency but speculate that it might arise either due to QI or as a conventional Onsager frequency corresponding to a FS that deviates from their DFT prediction. 
While the latter scenario seems more unlikely for a frequency that is absent in the magnetic torque and so strong in the SdH effect, QI can not be ruled out at this stage. However, the approximate frequency difference and mass difference may also be indicative of QPLOs. A more detailed analysis may be possible by investigating the FS and possible QI pathways of charge carriers in more detail.


\begin{table*}
    \centering
    \begin{tabular}{|l|l|}
    \hline
         \textbf{Material class} & \textbf{Materials and references} \\ \hline
         Topological semimetals & CoSi \cite{huber2023quantum}, MoSi$_2$ \cite{pavlosiuk2022giant}, WSi$_2$ \cite{pavlosiuk2022giant}, 
         \\ \hline
         Kagome materials & Co$_3$Sn$_2$S$_2$ \cite{ding2021quantum}
         \\ \hline
         Nodal line semimetals     & HfSiS \cite{vandelft2018electronhole}, ZrSiS \cite{muller2020determination}
          \\ \hline
         Weyl semimetals & WTe$_2$ \cite{zhu2015quantum,pan2022ultrahigh, wu2017threedimensionality,cai2015drastic}, PrAlSi \cite{wu2023fieldinduced}
         \\ \hline
         Oxychalcogenides &   KV$_2$Se$_2$O \cite{bai2024absence}
         \\ \hline
         Iron-based superconductors & FeSe$_{1-x}$S$_x$ \cite{reiss2020quenched,coldea2019evolution,leeb2024interband}
         \\ \hline
         Spin triplet superconductors & UTe$_2$ \cite{broyles2023revealing}
         \\ \hline
         Heavy-fermion compounds & 
         URu$_2$Si$_2$ \cite{hassinger2010similarity}
         \\ \hline
         Layered organic conductors & $\beta''$-(ET)$_4$(H$_3$O)[Fe(C$_2$O$_4$)$_3$]$\cdot$C$_6$H$_4$Cl$_2$ and others \cite{audouard2013organic}
         \\ \hline
         Skyrmion-hosting magnets & GdRu$_2$Si$_2$ \cite{matsuyama2023quantum} 
         \\ \hline
    \end{tabular}
    \caption{Selection of candidate materials and corresponding publications in which QPLOs may have been observed. The list is not meant to be exhaustive. The large variety of different material classes suggests that apparent quantum oscillation frequencies due to the QPLO mechanism may indeed be abundant in real materials and in the published literature. } 
    \label{tab:material_overview}
\end{table*}

\section{Conclusion}\label{sec:conclusion}
We have reviewed QO mechanisms that defy the semi-classical Onsager relation between the area of closed extremal orbits and the corresponding QO frequency. As a point of reference, a brief introduction to, both, the quantum mechanical and the semiclassical picture of conventional QOs was given. The novel mechanism of quasiparticle lifetime oscillations (QPLOs) was compared to other, well-known, non-Onsager QOs, namely magnetic breakdown (MB), magnetic interaction (MI), chemical potential oscillations (CPOs) and Stark quantum interference (QI), see table\,\ref{tab:mechanism_overview} for an overview. 
 
The objective of this work is to provide a field guide for identifying the origin of QO frequencies and distinguish non-Onsager QOs from conventional QOs that correspond to an extremal FS cross section. The ability to reliably distinguish Onsager and non-Onsager QOs is crucial for the correct identification of a material's electronic structure. Generally speaking, inexplicable frequencies have sometimes been attributed to ad-hoc FS reconstructions without explicitly excluding a non-Onsager origin. Vice versa, unexpected frequencies have been attributed to one of the above non-Onsager mechanisms without verifying whether the conditions for its occurrence were met.

This is in particular true for cases where frequencies have been attributed to QI without verifying the rather stringent conditions for its existence, i.e., partial MB at MB junctions and the existence of an effective interferometer set-up in the FS. We argue that some of these instances may, in fact, be better explained by QPLOs. {\cblue The interorbit scattering required for the occurrence of QPLOs can originate from impurity scattering~\cite{Leeb2021theory} but also from inter-orbit scattering by a dynamical boson, for example, phonons or spin fluctuations~\cite{mangeolle2024anomalous}. The latter can have  a non-monotonous temperature dependence because dynamical boson-induced QPLOs set in, once the dynamical boson is thermally activated. While understanding the microscopic origin is expected to differ from compound to compound, it allows in turn for a precise characterization of the scattering mechanisms~\cite{leeb2024interband}.}

We listed numerous materials belonging to various material classes (see table\,\ref{tab:material_overview}), where the available data suggests that QPLO may have already been observed. For some of these materials (section \ref{mat:CoSi}--\ref{mat:ZrSiS}), the published data allow for identification as QPLOs. We hope that this review motivates further research into QPLOs and increases the awareness for possible non-Onsager origins of QOs in general. 

QO research is a mature field of condensed matter physics with a history now spanning almost a century. Nevertheless, it continues to provide surprises both in theory and experiment. Originally, the observation of standard QOs was a surprise in itself, but, following the seminal Onsager-LK theory, QO research matured into the gold standard technique for identifying Fermi surface geometry and properties like the effective mass and Dingle temperature. Similarly, we hope that in the future, the observation of QPLOs will allow us to learn about otherwise hard-to-measure material properties, including different forms of interband scattering and different types of impurity contributions. Luckily, there seem to exist candidate materials aplenty.

\clearpage
\pagebreak

\begin{acknowledgments}
V.~L. acknowledges support from the Studienstiftung des deutschen Volkes. N.~H. and V.~L. acknowledge support from the TUM Graduate School. M.~A.~W. and C.~P. acknowledge support by the Deutsche Forschungsgemeinschaft (DFG, German Research Foundation) through TRR360-492547816 (ConQuMat) and DFG-GACR project WI3320/3 (project ID No. 323760292). C.P. acknowledges support via SPP 2137 (Skyrmionics) under grant no. PF393/19 (project ID  No. 403191981), DFG cluster EXC-2111 (project ID No. 390814868), and ERC Advanced Grant no. 788031 (ExQuiSid). J.~K. acknowledges support from the Imperial-TUM flagship partnership. J.~K. acknowledges support from the Deutsche Forschungsgemeinschaft (DFG, German Research Foundation) under Germany’s Excellence Strategy–EXC– 2111–390814868, DFG grants No. KN1254/1- 2, KN1254/2- 1, and TRR 360 - 492547816. The research is part of the Munich Quantum Valley, which is supported by the Bavarian state government with funds from the Hightech Agenda Bayern Plus.
\end{acknowledgments}

\clearpage
\bibliography{bib_zotero,bib_general,bib_CoSi_DiffFreq}

\clearpage
\tableofcontents

\end{document}